\documentstyle[aps,eqsecnum,epsfig]{revtex}
\begin{document}
\draft 
\title{Dynamics of structural models with a long-range interaction :
  glassy versus non-glassy behavior}

\author{V.G. Rostiashvili$^{(1,2)}$ and T.A. Vilgis$^{(1,3)}$}
\address{$^{(1)}$Max-Planck-Institut f\"ur Polymerforschung, Postfach 3148,
  D-55021 Mainz, Germany} \address{$^{(2)}$ Institute of Chemical Physics,
  Russian Academy of Science, 142432, Chernogolovka, Moscow region, Russia}
\address{$^{(3)}$ Laboratoire Europ{\'e}en Associ{\'e}, Insitut Charles
  Sadron. 6, rue Boussingault, F-67083 Strasbourg, France}

\date{\today} 

\maketitle

\begin{abstract}
  By making use of the Langevin dynamics and its generating functional (GF)
  formulation the influence of the long-range nature of the interaction on the
  tendency of the glass formation is systematically investigated. In doing so
  two types of models is considered: (i) the non-disordered model with a pure
  repulsive type of interaction and (ii) the model with a randomly distributed
  strength of interaction (a quenched disordered model). The long-ranged
  potential of interaction is scaled with a number of particles $N$ in such a
  way as to enable for GF the saddle-point treatment as well as the systematic
  $1/N$ - expansion around it. We show that the non-disordered model has no
  glass transition which is in line with the mean-field limit of the mode -
  coupling theory (MCT) predictions. On the other hand the model with a
  long-range interaction which above that has a quenched disorder leads to MC
  - equations which are generic for the $p$ - spin glass model and polymeric
  manifold in a random media.
\end{abstract}

\section{Introduction}
The theoretical description of slow dynamics is a crucial point to elucidate
the nature of the glass transition in structural glass - forming liquids. One
of the commonly used approaches, mode-coupling theory (MCT), was from the very
beginning designed for the supercooled simple liquids \cite{1} , i.e. for the
non - disordered models (as opposed to the models which contain quenched
disorder naturally).  Later it has been proven that MC - equations become
exact for a number of spin - glass models \cite{2,3,4,5,5',6,7} as well as for
the polymeric manifold in a random media \cite{6,7,8,8',9} (i.e. for the
models with quenched disorder) provided that the number of variables
components goes to the infinite.

Actually applicability of the MC - equations has been substantially extended
to the case when the time translation invariance and the fluctuation -
dissipation theorem do not hold any more \cite{7}. This striking similarity
between the models with and without quenched disorder suggests that the
effective disordered potential (e.g., in a supercooled liquid) is in a sense
`` self - induced'' and the difference between such a ``self - induced
disorder'' and the  quenched disorder might not be crucial \cite{6,7}.

In order to provide some insight into self - induced disorder we employed in
ref.\cite{10} a Feynman variational principle (VP) for a set of interacting
particles. Indeed it was shown that the VP is capable to treat metastable
states of the glass - forming system. The main point in ref.  \cite{10} was
that the partition function representation in terms of functional integrals is
twofold : (a) either as an integral over the local density , $\rho ({\bf r})$,
or (b) over the conjugated to $\rho({\bf r})$ field $\psi({\bf r})$. It has been
{\it assumed} that the {\it component average} free energy , $\bar F$, (which
is only meaningful in the supercooled regime) is equal to the variational free
energy $F_{{\rm VP}}$. There are at least four strong reasons in favor of this 
(at first sight not obvious) assumption.
\begin{itemize} 
\item The variational free energy $F_{\rm VP}$ is a upper bound for the {\it
    canonical} free energy, i.e., $F_{\rm c} \le \bar F = F_{{\rm VP}}$ , as
  it should be since $F_{\rm c} = \bar F - T\Sigma$ , where the complexity
  $\Sigma \ge 0$ \cite{11}.
\item After implementation of VP the initial problem is reduced to a self -
  consistent random field Ginzburg - Landau model (RFGLM). Then, as
  was  shown
  previously, the corresponding field $\psi(\bf r)$ must be upgraded to  a {\it replicated} field $\psi_{a} ({\bf r})$, where $a =
  1, .. ,n$ (with the final limit $n \to 0$) and the  density field 
  $\rho({\bf r})$ plays the role of an ``external'' field. Note that here the
  density $\rho(\bf r)$ is Gaussian due to the use of the VP. Eventually the
  correlators of $\psi$ and $\rho$ - fields can be determined self -
  consistently.
\item The resulting replicated partition function for RFGLM has a typical form
  which may eventually lead to the replica symmetry breaking (RSB), structural
  glass transition and the ``self - induced `` disorder.
\item Finally, in the case of the long - range interaction the
  partition function  allows the expansion around the saddle point ,
  or mean - field (MF), solution. It is possible to show then that the {\it next
    to the mean - field approximation} and VP merge
  and both become exact, i. e. $\bar F = F_{\rm c}$ and the glassy phase
  does not appear.
\end{itemize}
Some evidence for this behavior was deduced from the results for the
particles on a $M$- dimensional hypersphere \cite{12} at large dimensions, $M \to
\infty$. 

The aim of this paper is to face the full dynamical problem for a non -
disorded model with a long range interaction. Using the expansion around the
saddle point solution we derive the full equation of motion for the time
dependent density - density correlator and show that a ``glassy'' solution
does not exist. Conversly, if we add a term describing quenched disorder, by
random distribution of the strength of the interaction potential, then the
resulting equations of motion for two time density correlation and response
functions fall in the same class as MC - equations which have been widely
discussed \cite{2,3,4,5,5',6,7,8,8',9}. This means that the ``self - induced''
disorder is not generic for the pure model with the long-range interaction and
vice versa on addition of a quenched disorder the phase space becomes very
rugged resulting in slow dynamical processes. 

The paper is organized as follows. In the next section \ref{II} we first introduce the theorotical model
without quenched disorder . Its dynamics is
discussed by using the functional integral technique. The saddle point
solution yields the mean field dynamcis. Expansions around the saddle point
yield one loop corrections. The Legendre transformation provides the possibilty
of the analysis of the full  dynamic correlation matrix. In the section \ref{III}
quenched disorder is introduced by a ``random bond model'' and a Gaussian
disorder. The corresponding generating functional (GF) is computed by the
self-consistent Hartree approximation, which results in  a set of coupled
Langevin equations, that are solved in their asymptotic regimes. More details
on the calculations are laid out in the corresponding appendices.

\section{The model without quenched disorder}
\label{II}

We start from a simple model system which consists of interacting particles.
To do so, let us consider a set of $N (\gg 1)$ particles in $d$ - dimensional
space interacting by a pair potential of the form
\begin{eqnarray}
V(r) = \left(\frac{\mu}{N}\right) \frac{\exp(-\kappa r)}{4\pi
  r^{\alpha}}\label{1}\; .
\end{eqnarray}
This is a typical example of a long-range potential with a characteristic
length $\kappa$ and a coupling constant $\mu /N$. The choice of this potential
is twofold.  It contains a cut off at a $\kappa^{-1}$ and allows thus to
control the range of the interaction. Moreover at small scales
$(r<\kappa^{-1})$ it consists of a typical power law decay with long range
character, if $0<\alpha<2$. Therefore the so chosen potential allows to keep
control on range and nature of the interaction, which will become essential
below.  To ensure extensivity of the total interaction energy we require that
the integral $\int d^{d}V({\bf r}) = {\cal{O}}(N^{0})$, i.e. it does not
depend on the number of particles $N$. As a result we have $\kappa \propto
N^{-1/(d - \alpha)}$.
The intermolecular potential (\ref{1}) has the form of the generalized 
Kac potential
\begin{eqnarray}
V({\bf r}) = \kappa^d f(\kappa {\bf r}) \; ,\label{Kac}
\end{eqnarray}
 which has been used for the rigorous treatment of the
van-der-Waals theory \cite{13}. In order to provide conditions for the 
expansion around a saddle point, carried out later  on (see below), 
we should  require  that the
length $\kappa^{-1}$ must be larger compared to the characteristic size of the
system, which scales naturally as $N^{1/d}$) at $N \to \infty$. As a
consequence we find the limits for the range parameter $\alpha$
\begin{eqnarray}
0 < \alpha < d \; .
\end{eqnarray}
Below we shall restrict our considerations to the case: $d = 3, \alpha = 1$,
and the strength of the interaction $\mu > 0$ (pure repulsion) without loss of
generalization in the main statements we are going to predict. Then the
Fourier transformation of the potential (\ref{1}) takes especially simple form
\begin{eqnarray}
V({\bf k}) = \left(\frac{\mu}{N}\right)\frac{1}{k^{2} + \kappa^{2}}\; , 
\label{3}
\end{eqnarray}
which allows accurate analytic calculations.  In the limit $N \to \infty$ we
have thus $ \kappa^{2} \propto N^{-1}$, but the relevant minimum wave vector is
$k_{{\rm min}}^{2} \propto N ^{-2/3}$ and thus $\kappa^{2}$ can be actually
neglected under the  integration over the whole $k$ - space. As a result we arrive
formally at a {\it one - component plasma model} (OCP) \cite{14} where the
electroneutrality is implicitly provided by a neutralizing background.

\subsection{The generating functional method}
In the following we set up the relevant equations of motion for the model
system. We restrict ourselves to the  Langevin dynamics, which can be
comfortably  formulated
 in terms of dynamic functionals which  allows the systematic
$1/N$ - expansion treatment.  The Langevin dynamics of $N$ particles
interacting via the potential (\ref{1}) (at $d = 3, \alpha = 1$ and $\mu > 0$)
is described by the equation of motion
\begin{eqnarray}
m_{0}\frac{\partial^2}{\partial t^2} {\bf r}^{(p)}(t) +
\gamma_{0}\frac{\partial}{\partial t} {\bf r}^{(p)}(t) - \frac{\mu}{N} 
\sum_{m=1}^N \nabla v({\bf r}^{(p)} - {\bf r}^{(m)}) = {\bf
  f}^{(p)}(t)\; , \label{4}
\end{eqnarray}
where $m_{0}$ and $\gamma_{0}$ are the mass and the friction
coefficient respectively, $p = 1, 2, ..., N$ and $v(r;\kappa) =
\exp(-\kappa r)/4\pi r$. The random force in eq.(\ref{4}) is Gaussian
with $\left< f_{i}^{(p)}(t)\right> = 0$ and the correlator
\begin{eqnarray}
\left< f_{i}^{(p)}(t)f_{j}^{(n)}(t')\right> =
  2T\gamma_{0}\delta_{pm}\delta_{ij}\delta(t - t') \; , \label{5}
\end{eqnarray}
where from now on we work in units where the Boltzmann constant ${\rm k_{B}}=
1$.

As was mentioned, it is more convenient to reformulate the Langevin problem
(\ref{4})-(\ref{5}) by using the celebrated Martin-Siggia-Rose generating
functional (GF) method \cite{15}.  The method was first applied for
the $\phi^{4}$ - model with the long-range interaction in \cite{16} 
and for the polymer melt dynamics in \cite{17,18}.
Despite the fact that the Langevin equation (\ref{4}) is of the second 
order it is possible to show that the Jacobian which appear under transformation 
to the functional variables, still equal to one (see Appendix in
\cite{19}). After using this technique for the problem (\ref{4}) -
(\ref{5}), GF takes the form
\begin{eqnarray}
Z\{...\} = \int \prod_{p=1}^{N}D{\bf r}^{(p)}(t)D{\bf \hat
  r}^{(p)}(t)\exp \left\{\sum_{p=1}^{N}A_{0}[{\bf r}^{(p)},{\bf \hat
    r}^{(p)}] + \int dt \sum_{p=1}^{N}\sum_{m=1}^{N} \frac{\mu}{N}
  i{\hat r}_{j}^{(p)}(t) \nabla_{j}^{(p)}v\left({\bf r}^{(p)} - {\bf
      r}^{(m)}\right)\right\} \; , \label{6}
\end{eqnarray}
where the action of the free system
\begin{eqnarray}
A_{0}[{\bf r}^{(p)},{\bf \hat r}^{(p)}] = \int dt \left\{ T\gamma_{0}
  \left[i{\hat r}_{j}^{(p)}(t)\right]^{2} + i{\hat
    r}_{j}^{(p)}(t)\left[m_{0}\frac{\partial^2}{\partial t^2}
    {r}_{j}^{(p)}(t) +  \gamma_{0}\frac{\partial}{\partial t}
    {r}_{j}^{(p)}(t)  \right]\right\} \; . \label{7}
\end{eqnarray}
In the following we are going to transform this functional to collective
density variables. 
By using the transformations to the mass density
\begin{eqnarray}
\rho ({\bf r}) = \sum_{p=1}^{N} \delta ({\bf r} - {\bf r}^{(p)}(t))\label{8}
\end{eqnarray} 
and the longitudinal projection of the response field density
\begin{eqnarray}
\pi ({\bf r}) =   \sum_{p=1}^{N}i \hat r_{i}^{(p)}(t)\nabla_{i} \delta ({\bf r} - {\bf r}^{(p)}(t))\label{9} 
\end{eqnarray} 
for the GF one gets
\begin{eqnarray}
Z\left\{\chi_{\alpha}\right\}=\int \prod_{\alpha=0}^{1}
D\rho_{\alpha}(1)\exp\Bigg\{W\{\rho_{\alpha}\}-\frac{1}{2}\int d1d2\rho_{\alpha}(1)U_{\alpha\beta}(1,2)\rho_{\beta}(2)
+\int d1\rho_{\alpha}(1)\chi_{\alpha}(1)\Bigg\} \; ,\label{10}
\end{eqnarray}
where the summation over the repeated Greek indices is implied.
In eq.(\ref{10}) we have introduced 2-dimensional field
\begin{eqnarray}
\rho_{\alpha}(1) \equiv {\rho (1)\choose\pi (1)}\; , \label{11}
\end{eqnarray}
where $\alpha = 0,1$ and $1 \equiv ({\bf r}, t)$. The
``entropy'' of the free system is given as usual by
\begin{eqnarray}
W\{\rho,\pi\}&=&\log \int\prod_{p=1}^{N}D{\bf r}^{(p)}(t)D{\hat {\bf
    r}}^{(p)}(t)\exp\left\{\sum_{p=1}^{N} A_{0}\{{\bf r}^{(p)},{\hat {\bf
      r}}^{(p)}\}\right\}\nonumber\\
&\times&\delta\left[\rho({\bf r},t)-\sum_{p=1}^{N}\:\delta({\bf r}-{\bf
  r}^{(p)}(t))\right]\nonumber\\
&\times&\delta\left[\pi({\bf r},t)-\sum_{p=1}^{N}\:i{\hat r}_{j}^{(p)}(t)\nabla_{j}\delta({\bf r}-{\bf
  r}^{(p)}(t))\right] \; , \label{12}
\end{eqnarray}
and $U_{\alpha \beta}$ is the 2$\times$ 2 - interaction matrix
\begin{eqnarray}
U_{\alpha\beta}(1,2)=\left({0\atop V(|{\bf r}_{1}-{\bf r}_{2}|)}{V(|{\bf
    r}_{1}-{\bf r}_{2}|)\atop 0}\right)\label{13}
\end{eqnarray}
and $\chi_{\alpha}(1)$ is a source field.

An alternative valuable representation of GF can be obtained through the
``functional Fourier transformation'' 
\begin{eqnarray}
\exp\left\{F\{\psi_{\alpha}\}\right\} = \int
D\rho_{\alpha}(1)\exp\Bigg\{W\{\rho_{\alpha}\} - i\int d1\rho_{\alpha}(1)\psi_{\alpha}(1)\Bigg\}\label{14}
\end{eqnarray}
and its inversion
\begin{eqnarray}
\exp\left\{W\{\rho_{\alpha}\}\right\} = \int
D\psi_{\alpha}(1)\exp\Bigg\{F\{\psi_{\alpha}\} +i\int
d1\rho_{\alpha}(1)\psi_{\alpha}(1)\Bigg\} \; .\label{15}
\end{eqnarray}
The substitution of eq.(\ref{12}) into eq.(\ref{14}) leads to the
explicit expression for the free-system~GF
\begin{eqnarray}
\exp\left\{F\{\psi_{\alpha}\}\right\} &=& \int \prod_{p=1}^{N}D{\bf r}^{(p)}(t)D{\bf \hat
  r}^{(p)}(t)\exp \Bigg\{\sum_{p=1}^{N}A_{0}[{\bf r}^{(p)},{\bf \hat
    r}^{(p)}]\nonumber\\
 &-& i\sum_{p=1}^{N}\int dt \psi \left({\bf r}^{(p)}\right)
  +\left. i\sum_{p=1}^{N}\int dt i \hat r_{j}^{(p)}(t)\nabla_{j}\phi ({\bf
    r})\right|_{{\bf r}= {\bf r}^{(p)}(t)} \Bigg\}\; ,\label{16}
\end{eqnarray}
where $\psi (1)$ and $\phi (1)$ are components of the column - variable
\begin{eqnarray}
\psi_{\alpha}(1) \equiv {\psi (1)\choose\phi (1)} \; .\label{17}
\end{eqnarray}
By making use (\ref{15}) in (\ref{10}) and after functional
integration over $\rho_{\alpha}(1)$ one gets
\begin{eqnarray}
Z\left\{\chi_{\alpha}, \lambda_{\alpha}\right\}&=&\int \prod_{\alpha=0}^{1}
D\psi_{\alpha}(1)\exp\Bigg\{F\{\psi_{\alpha}\}+\nonumber\\
\frac{1}{2}\int
d1d2\bigl[i\psi_{\alpha}(1)&+&\chi_{\alpha}(1)\bigl]\bigl[U^{-1}\bigl]_{\alpha\beta}(
1,2)\bigl[i\psi_{\beta}(2)+ \chi_{\beta}(2)\bigl]
+\int d1\psi_{\alpha}(1)\lambda_{\alpha}(1)\Bigg\}\; ,\label{18}
\end{eqnarray}
where we have also add a source field $\lambda_{\alpha}(1)$ conjugated 
to  $\psi_{\alpha}(1)$. As a result eqs.(\ref{10}) and (\ref{18})
provide two equivalent representations of GF. For the purpose of expansion around the saddle point we use
representation (\ref{18}) at $\lambda_{\alpha}(1) = 0$ which after the 
transformation $\psi_{\alpha} \longrightarrow \psi_{\alpha} +
i\chi_{\alpha}$, yields
\begin{eqnarray}
Z\left\{\chi_{\alpha}\right\}=\int
\prod_{\alpha=0}^{1}D\psi_{\alpha}(1)\exp\Bigg\{ - N A\left[\psi_{\alpha};
  \chi_{\alpha}\right]\Bigg\}\; ,
\label{19}
\end{eqnarray}
which is appropriate for a saddle point integration, since the particle number
$N$ is large. 
The action hereby is given as
\begin{eqnarray}
A\left[\psi_{\alpha};\chi_{\alpha}\right]&=& \frac{1}{2}\int dt \int
d^3{\bf r}d^3{\bf r'}\psi_{\alpha}({\bf
  r},t)\left[v^{-1}\right]_{\alpha\beta}({\bf r}- {\bf r'}; \kappa
)\psi_{\beta}({\bf r'}, t) - 
\frac{1}{N}\log \int
\prod_{p=1}^{N}D{\bf r}^{(p)}(t)D{\bf \hat r}^{(p)}(t)\nonumber\\
&\times&\exp
\Bigg\{\sum_{p=1}^{N}A_{0}[{\bf r}^{(p)},{\bf \hat r}^{(p)}] -
i\sum_{p=1}^{N}\int dt
r_{\alpha}^{(p)}(t)\left[\psi_{\alpha}\left({\bf r}^{(p)}(t)\right) + i\chi_{\alpha}\left({\bf r}^{(p)}(t)\right)\right]\Bigg\}\; ,\label{20}
\end{eqnarray}
and the interaction matrix
\begin{eqnarray}
v_{\alpha\beta}({\bf r};\kappa)=\left({0\atop 1}{1\atop
    0}\right)\frac{\exp(-\kappa r)}{4\pi r} \; .
\label{21}
\end{eqnarray}
Recall that  the relation  $\kappa \propto N^{-1/2}$ is necessary  for the validity
of the saddle point integration. Moreover we have defined the column-vector
\begin{eqnarray}
r_{\alpha}^{(p)}(t) = {1\choose -i{\hat r}_{j}^{(p)}(t)\int
    d\tau\frac{\delta}{\delta r_{j}^{(p)}(\tau)}}\; 
\label{22}
\end{eqnarray}
for convenience.
\subsection{The saddle point solution and expansion
  around the SP}
Minimization of $A\left[\psi_{\alpha};\chi_{\alpha}\right]$ with
respect to $\psi_{\alpha}(1)$ leads to the SP - equations for the mean 
fields $\overline{\psi_{\alpha}}(1)$
\begin{eqnarray}
\overline{\psi_{\alpha}}({\bf r}, t)  =
-\frac{i\mu}{N}\int d^3{\bf r'}v_{\alpha\beta}({\bf r} - {\bf
  r'})\left<\rho_{\beta}({\bf r'},t)\right>_{{\rm SP}}\; ,\label{23}
\end{eqnarray}
where the average $\left< ... \right>_{{SP}}$ is calculated by using
the {\it cumulant} GF
\begin{eqnarray}
P_{{\rm SP}}\left\{\overline{\psi_{\alpha}} + i\chi_{\alpha}\right\} \equiv \frac{1}{N}\log \int
\prod_{p=1}^{N}D{\bf r}^{(p)}(t)D{\bf \hat r}^{(p)}(t)\exp
\Bigg\{\sum_{p=1}^{N}A_{0}[{\bf r}^{(p)},{\bf \hat r}^{(p)}] -\nonumber\\
i\sum_{p=1}^{N}\int dt
r_{\alpha}^{(p)}(t)\left[\overline{\psi_{\alpha}}\left({\bf r}^{(p)}(t)\right) + i\chi_{\alpha}\left({\bf r}^{(p)}(t)\right)\right]\Bigg\}\; .\label{24}
\end{eqnarray}
The correlation matrix in the random phase approximation (RPA) is
defined in such a way 
\begin{eqnarray}
S_{\alpha\beta}(1, 2) = \lim_{\overline{\psi_{\alpha}}+ i\chi_{\alpha} 
  \to 0}\left[ \frac{\delta \left< \rho_{\alpha}(1)\right>_{{\rm SP}}}{N
      \delta\chi_{\beta}(2)}\right]\; .
\label{25}
\end{eqnarray}
After linearization of eq.(\ref{24}) with respect to
$\overline{\psi_{\alpha}}+ i\chi_{\alpha}$ the 2$\times$2 - RPA
correlation matrix is easily found to coincide with the well
known form \cite{17} 
\begin{eqnarray}
S_{\alpha\beta}(1, 2)= \left\{\left[\hat F^{-1} + \mu \hat v
  \right]^{-1}\right\}_{\alpha\beta}(1, 2)\; ,
\label{26}
\end{eqnarray}
where $\hat v$ is the interaction matrix (\ref{21}) and $F_{\alpha \beta}$ is
the correlation matrix for the {\it free system} $F_{\alpha\beta}(1, 2) =
\left< \Delta\rho_{\alpha}(1)\Delta\rho_{\beta}(2)\right>_{0}/N$ has the form
\begin{eqnarray}
F_{\alpha\beta}(1,2)=\left({F_{00}(1, 2)
\atop F_{10}(1, 2)}{F_{01}(1, 2) \atop 0}\right)\; .
\label{27}
\end{eqnarray}
In eq.(\ref{27}) $F_{01}(1, 2)$ and $F_{10}(1, 2)$ are response
functions whereas $F_{00}(1, 2)$ stands for the correlation
function. The relation between them is given by the fluctuation
dissipation theorem (FDT) which in $({\bf k}, t)$- representation has
the form
\begin{eqnarray}
-\beta\frac{\partial}{\partial t}F_{00}({\bf k},t)=F_{01}({\bf
  k},t)-F_{10}({\bf k},t)\; .\label{28}
\end{eqnarray}
It is easy to check that in this case the FDT for the RPA type correlation
matrix (\ref{26}) also holds
\begin{eqnarray}
-\beta\frac{\partial}{\partial t}S_{00}({\bf k},t)=S_{01}({\bf
  k},t)-S_{10}({\bf k},t)\; ,
\label{29}
\end{eqnarray}
where $\beta = 1/T$ is the inverse temperature.  The corresponding elements of
the RPA - matrix (\ref{26}) are of an especially simple form in the Fourier -
$({\bf k},\omega)$ - representation, namely
\begin{eqnarray}
S_{00}({\bf k},\omega)&=&\frac{F_{00}({\bf
    k},\omega)}{\left[1+v(k)F_{10}({\bf
      k},\omega)\right]\left[1+v(k)F_{01}({\bf k},\omega)\right]}
\label{RPAres1}\\
S_{01}({\bf k},\omega)&=&\frac{F_{01}({\bf
    k},\omega)}{1+v(k)F_{01}({\bf k},\omega)}
\label{RPAres2}\\
S_{10}({\bf k},\omega)&=&\frac{F_{10}({\bf
    k},\omega)}{1+v(k)F_{10}({\bf k},\omega)}\; .
\label{RPAres3}
\end{eqnarray}
It turns out interesting to recover the wellknown form in the static limit,
where we have  $S_{01}({\bf k},\omega \to 0 ) = \beta S_{{\rm st}}({\bf k}) =
\left[\left(\beta F_{{\rm st}}\right)^{-1} + \mu k^{-2}\right]$ and
for the correlator $S_{{\rm RPA}}({\bf k}) = S_{{\rm st}}({\bf
  k})/\rho_{0}$ one gets
\begin{eqnarray}
S_{{\rm RPA}}({\bf k}) = \frac{1}{ 1 + \frac{\beta\mu\rho_{0}}{k^{2}}}\; ,\label{30}
\end{eqnarray}
where we have used $F_{{\rm st}} = \rho_{0}$ .
This expression is completely equivalent to the  correlator for the
OCP-model (see eq.(10.1.7) in \cite{14}) with the direct correlation
function $c({\bf k}) = -\mu\beta/k^{2}$ and the Debye wavenumber
$k_{{\rm D}} = \left(\beta\mu\rho_{0}\right)^{1/2}$.

Now let us expand the action (\ref{20}) around SP-solution (\ref{23})
up to the second order with respect to the fluctuations
$\psi_{\alpha}(1) - \overline{\psi_{\alpha}}(1)$. After the functional 
integration we arrive at the following result for the GF
\begin{eqnarray}
P\left\{\chi_{\alpha}\right\} &\equiv& \frac{1}{N} \log
Z\left\{\chi_{\alpha}\right\}\nonumber\\
&=& P_{{\rm SP}}\left\{\overline{\psi_{\alpha}} +  \chi_{\alpha}\right\} 
- \frac{1}{2N}{\rm Tr} \left[\log T_{\alpha\beta}(1, 2)\right]\; ,\label{31}
\end{eqnarray}
where $T_{\alpha \beta}(1,2)$ is the inverse matrix of the effective
interactions \cite{effpot}
\begin{eqnarray}
T_{\alpha\beta}(1, 2) =
\frac{1}{\mu}\left[v^{-1}\right]_{\alpha\beta}(1, 2) +
  \frac{1}{N}\left<\Delta\rho_{\alpha}(1)\Delta\rho_{\beta}(2)\right>_{{\rm SP}}\; .\label{32}
\end{eqnarray}
In eqs.(\ref{31}) - (\ref{32}) we deliberately keep external field
$\chi_{\alpha}(1)$ nonzero because it to be used in the next
subsection for the Legendre transformation.
\subsection{The Legendre transformation}
The functional Legendre transformation is a general way to provide the 
Dyson equation for the {\it full correlation matrix}
$G_{\alpha\beta}(1, 2)$ \cite{20}. In doing so the {\it irreducible}
GF, $\Gamma\left\{\left<\rho_{\alpha}(1)\right>\right\}$ , is defined by
the identity
\begin{eqnarray}
\Gamma\left\{\left<\rho_{\alpha}(1)\right>\right\} +
P\left\{\left<\chi_{\alpha}(1)\right>\right\} = \int d1
\left<\rho_{\alpha}(1)\right>\chi_{\alpha}(1)\; .\label{33} 
\end{eqnarray}
By doing functional differentiation of (\ref{33}) one gets
\begin{eqnarray}
\chi_{\alpha}(1) = \frac{\delta\Gamma\left\{\left<\rho_{\alpha}(1)\right>\right\}}{\delta\left<\rho_{\alpha}(1)\right>}\label{34}
\end{eqnarray}
and
\begin{eqnarray}
\left[G^{-1}\right]_{\alpha\beta}(1, 2) =
\frac{\delta^{2}\Gamma\left\{\left<\rho_{\alpha}(1)\right>\right\}}{\delta\left<\rho_{\alpha}(1)\right> \delta\left<\rho_{\beta}(2)\right> }\; .\label{35}
\end{eqnarray}
Taking into account the result in 
eq.(\ref{31}) we find the following result for GF
\begin{eqnarray}
\Gamma\left\{\left<\rho_{\alpha}(1)\right>\right\} = \Gamma_{{\rm
    SP}}\left\{\left<\rho_{\alpha}(1)\right>\right\} +\frac{1}{2N}{\rm Tr} \left[\log T_{\alpha\beta}(1, 2)\right]\; ,\label{35'}
\end{eqnarray}
where 
\begin{eqnarray}
\Gamma_{{\rm
    SP}}\left\{\left<\rho_{\alpha}(1)\right>\right\} = - P_{{\rm
    SP}}\left\{\chi_{\alpha}\right\} + \int d1
\left<\rho_{\alpha}(1)\right>\chi_{\alpha}(1)\; .\label{36} 
\end{eqnarray}
In eq.(\ref{35'}) one should consider $\chi_{\alpha}(1)$ as a functional of
$\left<\rho_{\alpha}(1)\right>$ given by eq.(\ref{34}). Double differentiation
of eq.(\ref{35'}) leads to an equation of the Dyson form
\begin{eqnarray}
\left[G^{-1}\right]_{\alpha\beta}(1, 2) =
\left[S^{-1}\right]_{\alpha\beta}(1, 2) - \Sigma_{\alpha\beta}(1, 2)\; ,\label{37} 
\end{eqnarray}
where the RPA - correlation matrix, $S_{\alpha\beta}(1, 2)$,  is defined by eqs.(\ref{RPAres1}) -
(\ref{RPAres3}) and the ``self - energy'' functional
$\Sigma_{\alpha\beta}(1, 2)$ has the form
\begin{eqnarray}
\Sigma_{\alpha\beta}(1, 2) = - \frac{1}{2N}{\rm Tr} \left\{\frac{\delta^{2}}{\delta\left<\rho_{\alpha}(1)\right> \delta\left<\rho_{\beta}(2)\right>}  \log T_{\gamma\delta}(3, 4)\right\}_{\chi_{\alpha}=0}\; .\label{38}
\end{eqnarray}
In eq.(\ref{38}) the ``trace'' is taken over the variables 3,4 and indices
$\gamma, \delta$. The explicit differentiation in eq.(\ref{38}) leads
to the result
\begin{eqnarray}
\Sigma_{\alpha\beta}(1, 2) = - \frac{1}{2N}{\rm Tr}
\left\{\hat{T}^{-1}\frac{\delta^{2}\hat{T}}{\delta\left<\rho_{\alpha}(1)\right> \delta\left<\rho_{\beta}(2)\right>} \right\}_{\chi_{\alpha}=0}\; ,\label{39}
\end{eqnarray}
where $\hat{T}$ is a short hand notation of the matrix
$T_{\gamma\delta}(3, 4)$ and we have took into account that $\delta
T_{\alpha\beta}(1, 2)/ \delta\chi_{\gamma}(3) =
\left<\Delta\rho_{\alpha}(1)
  \Delta\rho_{\beta}(2)\Delta\rho_{\gamma}(3)\right>_{{\rm SP}}/N = 0$ 
at $\chi_{\alpha}=0$ because the fluctuations are Gaussian. Further
calculation yields
\begin{eqnarray}
\frac{\delta^{2}T_{\gamma\delta}(3,
  4)}{\delta\left<\rho_{\alpha}(1)\right>
  \delta\left<\rho_{\beta}(2)\right>} = \int d5d6 \frac{1}{N}\left<\Delta\rho_{\gamma}(3)
  \Delta\rho_{\delta}(4)\Delta\rho_{\omega}(5)\Delta\rho_{\chi}(6)\right>_{{\rm SP}}R_{\omega\beta}(5, 2) R_{\chi\alpha}(6, 1)\;,\label{40}
\end{eqnarray}
where
\begin{eqnarray}
R_{\alpha\beta}(1, 2) = \frac{\delta\vartheta_{\alpha}(1)}{\delta\left<\rho_{\beta}(2)\right>}\label{41}
\end{eqnarray}
and the full mean field
\begin{eqnarray}
\vartheta_{\alpha}(1) = -i \overline{\psi_{\alpha}}(1) + \chi_{\alpha}(1)\; .\label{42}
\end{eqnarray}
The expression for $R_{\alpha\beta}(1, 2)$ can be easily found by
differentiation of eq.(\ref{42}) with respect to
$\left<\rho_{\beta}(2)\right>$. Taking into account eqs.(\ref{23}), (\ref{34}) and
(\ref{35}) at $\chi_{\alpha} \to 0$ one gets
\begin{eqnarray}
R_{\alpha\beta}(1, 2) =\left[G^{-1}\right]_{\alpha\beta}(1, 2) - \mu
\int d4 d3 v_{\alpha\omega}(1, 4) S_{\omega\gamma}(4,
3)R_{\gamma\beta}(3, 2)\label{43}
\end{eqnarray}
or finally
\begin{eqnarray}
R_{\alpha\beta}(1, 2) = \int d3 \left\{\left[\hat 1 +\mu \hat v \hat S
  \right]^{-1}\right\}_{\alpha\gamma}(1, 3)\left[\hat G^{-1}\right]_{\gamma\beta}(3, 2)\; ,\label{44} 
\end{eqnarray}
where the hatted variables stands for the corresponding 2$\times$2
matrices. Substitution eqs.(\ref{44}) and (\ref{40}) in eq.(\ref{39}) yields
\begin{eqnarray}
\Sigma_{\alpha\beta}(1, 2) = - \int d3 d4 K_{\gamma\delta}(3, 4)\left[G^{-1}\right]_{\gamma\alpha}(3, 1)\left[G^{-1}\right]_{\delta\beta}(4, 2)\;,\label{45}
\end{eqnarray}
where 2$\times$2 vertex - matrix has the form
\begin{eqnarray}
 K_{\alpha\beta}(1, 2) &=& \int d3 d4 d5 d6 \left\{\left[(\mu \hat v)^{-1} + \hat S \right]^{-1}\right\}_{\delta\gamma}(4, 3)
S^{(4)}_{\gamma\delta\omega\chi}(3, 4, 5, 6)\nonumber\\
&\times&\left\{\left[\hat 1 +\mu \hat v \hat S
  \right]^{-1}\right\}_{\omega\alpha}(5, 1)\left\{\left[\hat 1 +\mu \hat v \hat S
  \right]^{-1}\right\}_{\chi\beta}(6, 2)\; .\label{46}
\end{eqnarray}
In eq.(\ref{46})  $ S^{(4)}_{\alpha\beta\gamma\delta}(1, 2, 3, 4)$ is
the 4 - point (response) correlator matrix in RPA
\begin{eqnarray}
 S^{(4)}_{\alpha\beta\gamma\delta}(1, 2, 3, 4) = \frac{1}{N^2}\left<\Delta\rho_{\alpha}(1)
  \Delta\rho_{\beta}(2)\Delta\rho_{\gamma}(3)\Delta\rho_{\delta}(4)\right>_{{\rm SP}}\; .\label{47}
\end{eqnarray}
The explicit calculation of $ S^{(4)}_{\alpha\beta\gamma\delta}(1, 2,
3, 4)$ is implemented in Appendix B. The vertex - matrix can be seen
as a one-loop diagram (see Fig. 1). 

\begin{center}
\begin{minipage}{12cm}
\begin{figure}
\epsfig{file=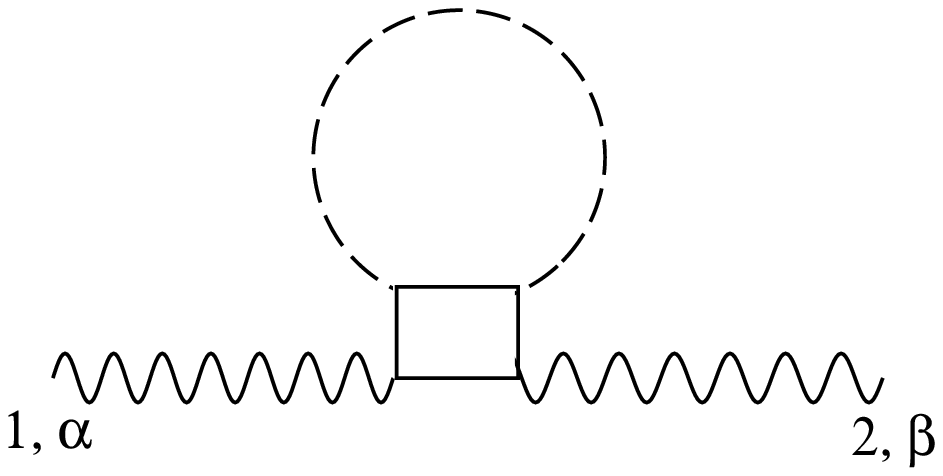,width=6cm}\\
{\caption 
{\footnotesize
Diagramatic interpretation of the vertex - matrix: the
  rectangle corresponds to
  $S_{\alpha\beta\gamma\delta}^{(4)}(1,2,3,4)$; the dash line - to the 
  effective interaction matrix $\left[(\mu {\hat v})^{-1} + {\hat
      S}\right]^{-1}$; the wave line - to $\left[ {\hat 1} + \mu {\hat 
      v}{\hat S}\right]^{-1}$.
} 
}
\end{figure}
\end{minipage}
\end{center}
The higher loops contributions
 which  include generally speaking 2m - point correlators
, $ S^{(2m)}_{\alpha\beta \ldots \gamma}(1,2,\ldots , 2m)$,  can be also considered, however the ``self -
energy'' has still the same convolution structure : $\hat \Sigma = \hat
G^{-1}\ast \tilde K \ast \hat G^{-1}$. Here the vertex - matrix
$\tilde K_{\alpha\beta}(1, 2)$ is calculated in RPA only . That is why 
these  contributions basically do not change our results.

As a result in the $({\bf k}, \omega)$- representation the Dyson
equation (\ref{37}) with the ``self - energy'' functional   (\ref{45})
reduces to   a quadratic
 one 
\begin{eqnarray}
G_{\alpha\gamma}({\bf k},
\omega)\left[S^{-1}\right]_{\gamma\delta}({\bf k},
\omega)G_{\delta\beta}({\bf k},\omega) - G_{\alpha\beta}({\bf k},
\omega) + K_{\alpha\beta}({\bf k}, \omega) = 0\; .\label{48}
\end{eqnarray}
The coefficients of the eq.(\ref{48}) trace the problem back to the {\it free
  system dynamics} which is embodied in the correlation matrices
$F_{\alpha\beta}(1, 2)$ and $F^{(4)}_{\alpha\beta\gamma\delta}(1, 2, 3, 4)$.
It is not surprising then that a specification of the model for the free
system dynamics is necessary, before going to the investigation of
eq.(\ref{48}).

\subsection{Analysis of the equation for the full  correlation matrix}

As we have mentioned the explicit solution of eq.(\ref{48}) needs the
specification of the free system dynamics. Two simple models are most
amenable for the theoretical treatment~ : free diffusion model (FDM) and  the
relaxation time approximation model (RTAM~) \cite{14,21}. The latter
provide more reasonable dynamical information also for short time
intervals, $\Delta t < m_{0}/\gamma_{0}$ , where the FDM completely
failed (e.g., the sum rule does not hold). It turns out that upon
calculation of the trace in eq.(\ref{46}) the integral is ultraviolet
- divergent for FDM and only RTAM leads to the finite result. The
matrix 
elements for RTAM has the form 
\begin{eqnarray}
F_{00}({\bf k}, \omega) &=& \frac{2 F_{{\rm st}} k^2 D}{\omega^2 + (k^2
  D - \omega^2 \tau_{0})^{2}}\\
F_{01}({\bf k}, \omega) &=& \frac{\beta F_{{\rm st}} k^2 D}{-i\omega + k^2
  D - \omega^2 \tau_{0} }\\
F_{10}({\bf k}, \omega) &=& \frac{\beta F_{{\rm st}} k^2 D}{i\omega + k^2
  D - \omega^2 \tau_{0} }\; ,\label{49}
\end{eqnarray}
where we introduced the diffusion coefficient $D = T/\gamma_{0}$, the
characteristic time scale $\tau_{0} = m_{0}/\gamma_{0}$, and $F_{{\rm st}} =
\rho_{0}$ for the overall density.  At $\tau_{0} = 0$ we return to FDM.  In
the case of RTAM the solution of eq.(\ref{48}) for the full correlation matrix
reads
\begin{eqnarray}
G_{01}({\bf k},\omega) &=& \frac{1 + \sqrt{1 - 4\left[-i\omega\tau_{\rm c}
      - \omega^{2}\tau_{0}\tau_{\rm c} + \chi_{st}^{-1}\right]K_{01}({\bf k},
    \omega)}}{2 \left[-i\omega\tau_{\rm c} -
    \omega^{2}\tau_{0}
\tau_{\rm c} + \chi_{st}^{-1}(k)\right]}\label{49'}\\
G_{10}({\bf k},\omega) &=& G_{01}(-{\bf k},-\omega) \label{50} 
\end{eqnarray}
and
\begin{eqnarray}
G_{00}({\bf k},\omega) = \frac{\frac{\tau_{\rm c}}{2
    \beta}\left|\frac{1 + \sqrt{1 - 4 \left[-i\omega\tau_{\rm c} -
          \omega^2 \tau_{0} \tau_{\rm c} + \chi_{st}^{-1}({\bf k})\right]
        K_{01}({\bf k},\omega)}}{- i\omega\tau_{\rm c} - \omega^2 \tau_{0} \tau_{\rm c} + \chi_{st}^{-1}({\bf k}) }\right|^2 - K_{00}({\bf k},
  \omega)}{{\rm Re} \left\{\sqrt{1 - 4\left[ - i \omega\tau_{\rm c} -\omega^2 \tau_{0} \tau_{\rm c} + \chi_{st}^{-1}({\bf k}) \right] K_{01}({\bf k},
  \omega)}\right\}}\; . \label{51}
\end{eqnarray}
The explicit calculation of the matrix $K_{\alpha\beta}({\bf k},\omega)$ (see
eq.(\ref{46}) is given in the Appendix C. The overall behavior of the
correlation function $G_{00}({\bf k},\omega)$ according eq. (\ref{51}) is
shown in Fig. 2 (at $\mu = 10,\; \beta = 0.1 ,\; \rho_{0} = 1\;
\mbox{and}\; \tau_{0} = 0.1$). It can be seen clearly that there no
singularity appears at $\omega \to 0$. 

\begin{center}
\begin{minipage}{12cm}
\begin{figure}
\epsfig{file=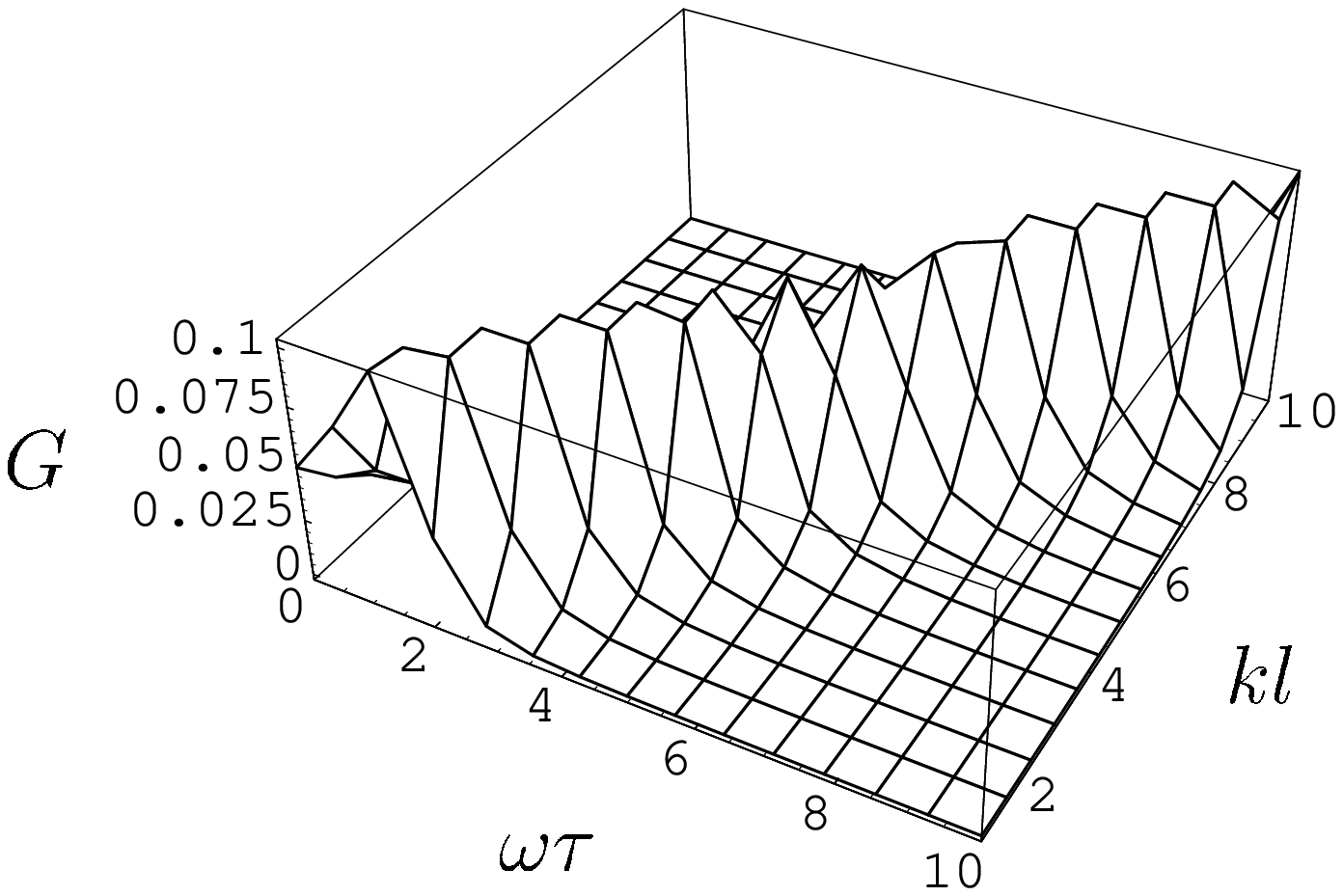,width=8cm}\\
{\caption 
{\footnotesize
The correlation function $G_{00}({\bf k},\omega)$ vs rescaled 
  variables $\omega\tau_{0}$  and  $kl_{0}$, where $\tau_{0} =
  m_{0}/\gamma_{0}$ and $l_{0} = (\tau_{0}/\beta\gamma_{0})^{1/2}$
} 
}
\end{figure}
\end{minipage}
\end{center}

The singularity, however, might be
responsible for a glass transition. Instead, the low frequency limit of
$G_{00}({\bf k},\omega)$ slowly changed with control parameters (which is not
shown in Fig.2).  That means that for the non - disordered model with a
general repulsive long ranged potential (\ref{1}) the glass transition is not
generic. This very important conclusion suggest that for the model with a long
range interaction the phase space is too smooth to show a glass
transition. In order to obtain a glass like transition  a 
competing interactions or a quenched disorder should be added. This
leads to glassy dynamics, as we will show in the next section.

It is interesting to note, that for the generalized Kac potential eq.
(\ref{Kac}), where $f({\bf r})$ and its Fourier transformation are positive
definite functions, MCT - memory kernel vanishes at $\kappa \to
0$\cite{Fuchs}. The corresponding argumentation is relegated to the 
Appendix D.  The explanation for this result  lies  in 
the fact that the ``cage effect'', which is a cornerstone of MCT, is
missing in the MF - limit.

The ``glass transition'' which has been studied in ref.\cite{Klein}
for the particles interacted via the Kac potential (\ref{Kac}) has a
completely different nature.
In ref.\cite{Klein} the function $f({\bf r})$ has a step form so that
its Fourier transform $f({\bf k})$ is negative at some value of ${\bf
  k}$ . As a result the system becomes unstable and a nonuniform
configuration where the particles are grouped into ``clumps'' shows
up. It was found that the slow dynamics of the MF - model is
associated with these clumps and does not touch a single particle
motion. Obviously it is different from the conventional glass
transition \cite{1}.

\section{The structural model with a competing quenched interactions}
\label{III}
\subsection{Specification of the model}
In the previous sections we have shown in detail that in the absence of
disorder the dynamical spectrum changed monotonically with a control parameter
and no glassy dynamics can be seen. The natural question which arises now is:
How will the introduction of competing interactions and / or quenched disorder
affect the dynamics of the system discussed above?  To provide an answer to
this question we will use already existing models 
of heteropolymers  and their disordered   two body interaction
\cite{22,23,24,25}. The use of these models and techniques are here natural,
since the behavior of heteropolymers is well discussed in the literature.
 In principle, two practical
possibilities exist.
\begin{itemize}
\item The strength of the  two- body interaction , $\mu$, in
  eq.(\ref{1}) is now a random function of all pairs of the interacted 
  particles, $\mu_{pm}$ (``random - bond model'').
\item Each particle carry a single ``charge'' $\sigma_{p}$, so that
  $\mu_{pm} = \xi_{0} + b\sigma_{p}\sigma_{m}$ and each $\sigma_{p}$
  is randomly distributed (``random sequence model''). 
\end{itemize}
It turns out to be sufficient for the purpose of this paper to restrict
ourself only with the ``random - bond model'' , where $\mu_{pm}$ does not
depend from the choice of pairs and has a Gaussian distribution
\begin{eqnarray}
P\{\mu_{pm}\} \propto \exp
    \left\{-\frac{(\mu_{pm} - \xi_{0})^2}{2\chi^{2}}\right\}\; .\label{52}
\end{eqnarray}
The competing long range interactions  frustrates the
system of particles and the question is whether  a glass transition
exist 
or not. Normally frustration and frozen disorder is enough for the existence
of glassy phases. Here the problem is more complicated, since the long range
nature of the interaction may provide opposite effects.

The averaging  over the quenched disorder in eq.(\ref{6}) (after the
substitution $\mu \to \mu_{pm})$ can be carried out just in the same way as
in ref.\cite{26,27}. Similarly, typical two-time dependent terms
immediately appear. They are also bilinear with respect to the forces
of interaction $\nabla_{j}v({\bf r})$ . In order to rationalize these
terms it is convenient to introduce (besides the mass density
(\ref{8}) and the response field density (\ref{9})) the following
collective variables
\begin{eqnarray}
Q_{0}({\bf r},t;{\bf r'}, t')&=& \sum_{p=1}^{N}\delta({\bf r} - {\bf
  r}^{p}(t))\delta({\bf r'} - {\bf r'}^{p}(t'))\nonumber \\
Q_{1}({\bf r},t;{\bf r'}, t')&=& -\sum_{p=1}^{N}i\hat
r_{j}(t)\nabla_{j}\delta({\bf r} - {\bf r}^{p}(t))i\hat r_{l}(t')\nabla_{l}\delta({\bf r'} - {\bf r'}^{p}(t'))\nonumber\\
Q_{2}({\bf r},t;{\bf r'}, t')&=& \sum_{p=1}^{N}i\hat
r_{j}(t)\nabla_{j}\delta({\bf r} - {\bf r}^{p}(t))\delta({\bf r'} - {\bf r'}^{p}(t'))\nonumber\\
Q_{3}({\bf r},t;{\bf r'}, t')&=& -\sum_{p=1}^{N}i\hat
r_{j}(t')\nabla_{j}\delta({\bf r'} - {\bf r}^{p}(t'))\delta({\bf r} - {\bf r}^{p}(t))\; .\label{53}
\end{eqnarray}
After the introduction of the  4-dimensional column - fields
\begin{eqnarray}
Q_{{\rm a}}(1; 1') = \left[\begin{array}{c}
Q_{0}(1; 1')\\Q_{1}(1; 1')\\Q_{2}(1; 1')\\Q_{3}(1; 1')\label{54}
\end{array}\right]\; ,
\end{eqnarray}
where  $a= 1, 2, 3, 4$and 4$\times$4 - matrix
\begin{eqnarray}
\Gamma_{{\rm ab}}(1, 2, 3, 4) = \left[\begin{array}{cccc}
0&v(1, 3)v(2, 4)&0&0\\v(1, 3)v(2, 4)&0&0&0\\0&0&0&v(1, 4)v(3, 2)\\0&0&v(1, 4)v(3, 2)&0
\end{array}\right]\label{55}
\end{eqnarray}
the whole expression for GF takes the form
\begin{eqnarray}
\left< Z \right>_{\rm av}\left\{\chi_{\alpha} , H_{a}\right\} &=&
\int\prod\limits_{\alpha=0}^{1}\prod\limits_{a =
  0}^{3}D\rho_{\alpha}(1)DQ_{\rm
  a}(1;1')\exp\Bigg\{\widetilde{W}\left\{\rho_{\alpha}(1);Q_{\rm
    a}(1;1')\right\}\nonumber\\
&-& \frac{\xi_{0}}{2N}\int d1d2\rho_{\alpha}(1)U_{\alpha\beta}(1,2)\rho_{\beta}(2)
+\int d1\rho_{\alpha}(1)\chi_{\alpha}(1)\nonumber\\
&-& \frac{\chi^2}{4N^2}\int d1 d2 d3 d4 Q_{\rm
    a}(1;2)\Gamma_{{\rm ab}}(1, 2, 3, 4) Q_{\rm
    b}(3;4) + \int d1 d2   Q_{\rm a}(1;2)H_{a}(1;2)\Bigg\}\; ,\label{56}
\end{eqnarray}
where the entropy is given by
\begin{eqnarray}
\widetilde{W}\left\{\rho_{\alpha}(1);Q_{\rm
    a}(1;1')\right\} &=& \log \int\prod_{p=1}^{N}D{\bf r}^{(p)}(t)D{\hat {\bf
    r}}^{(p)}(t)\exp\left\{A_{0}\{{\bf r}^{(p)},{\hat {\bf
      r}}^{(p)}\}\right\}\nonumber\\
&\times&\prod_{\alpha = 0}^{1}\delta\left[\rho_{\alpha}(1)-\sum_{p=1}^{N}\:r_{\alpha}^{(p)}(1)\delta({\bf r}_{1}-{\bf
  r}^{(p)}(t))\right]\nonumber\\
&\times&\prod_{{\rm a}=1}^{4}\delta\left[Q_{{\rm a}}(1;
  2)-\sum_{p=1}^{N}\:p_{{\rm a}}^{(p)}(1; 2)\delta({\bf r}_{1}-{\bf
  r}^{(p)}(t_{1}))\delta({\bf r}_{2}-{\bf
  r}^{(p)}(t_{2})) \right]\;  . \label{57}
\end{eqnarray}
We had used the column - operators
\begin{eqnarray}
r_{\alpha}^{(p)}(1) = \left(\begin{array}{c}
1\\i\hat r_{j}^{(p)}(t_{1})\nabla_{j, 1}
\end{array}\right)\; \; p_{{\rm a}}^{(p)}(1; 2) = \left(\begin{array}{c}1\\- i\hat r_{j}^{(p)}(t_{1})\nabla_{j, 1}i\hat r_{l}^{(p)}(t_{2})\nabla_{l, 2}
  \\i\hat r_{j}^{(p)}(t_{1})\nabla_{j, 1}  \\- i\hat r_{j}^{(p)}(t_{2})\nabla_{j, 2} \end{array}\right)\label{58}
\end{eqnarray}
and the external field, $H_{{\rm a}}(1; 2)$, conjugated to $Q_{{\rm
    a}}(1; 2)$,  has been introduced also.

The two- point collective fields (\ref{53}) have a meaning of the
dynamical ``overlaps''. It is  a dynamical generalization of the Parisi
``overlaps'' in a replica space \cite{28}. For example $Q_{0}(1; 1')$
quantify density - density and $Q_{2}(1; 1')$ response - density
overlaps respectively between two space-time points. The ``entropy''
(\ref{57}) corresponds to the volume in the dynamical phase space when
not only fields $\rho_{\alpha}(1)$ but also overlaps  $Q_{{\rm a}}(1;
  1')$ are given. In a sense the ``entropy'' (\ref{57}) is again the
  generalization of the entropy for the heteropolymer spanned in a
  replica space at the given set of ``overlaps'' \cite{24}.

\subsection{The saddle point treatment}
Let us introduce the functional $\widetilde{F}\left\{\psi_{\alpha}(1);\Phi_{\rm
    a}(1;1')\right\}$ by the functional Fourier transformation
\begin{eqnarray}
\exp\{\widetilde{W}\left\{\rho_{\alpha}(1);Q_{\rm
    a}(1;1')\right\}\} &=& \int\prod\limits_{\alpha=0}^{1}\prod\limits_{a =
  0}^{3}D\psi_{\alpha}(1)D\Phi_{\rm
  a}(1;1')\exp\Bigg\{\widetilde{F}\left\{\psi_{\alpha}(1);\Phi_{\rm
    a}(1;1')\right\}\nonumber\\
&+& i\int d1\rho_{\alpha}(1)\psi_{\alpha}(1)+ i\int d1 d2   Q_{\rm a}(1;2)\Phi_{a}(1;2)\Bigg\}\; .\label{59}
\end{eqnarray}
After substitution in eq.(\ref{56}) and integration over
$\rho_{\alpha}$ and $Q_{\rm a}(1;2)$ one gets
\begin{eqnarray}
\left<Z \right>_{\rm av}\left\{\chi_{\alpha} , H_{a}\right\} &=&
\int\prod\limits_{\alpha=0}^{1}\prod\limits_{a =
  0}^{3}D\psi_{\alpha}(1)D\Phi_{\rm
  a}(1;1')\exp\Bigg\{\widetilde{F}\left\{\psi_{\alpha}(1);\Phi_{\rm
    a}(1;1')\right\}\nonumber\\
&-& \frac{N}{2\xi_{0} }\int d1d2\psi_{\alpha}(1)\left[v^{-1}\right]_{\alpha\beta}(1,2)\psi_{\beta}(2)\nonumber\\
&-& \frac{N}{\chi_{0}^2}\int d1 d2 d3 d4 \Phi_{\rm
    a}(1;2)\left[\Gamma^{-1}\right]_{{\rm ab}}(1, 2, 3, 4) \Phi_{\rm
    b}(3;4)\Bigg\}\; ,\label{60}
\end{eqnarray}
where
\begin{eqnarray}
\widetilde{F}\left\{\psi_{\alpha}(1);\Phi_{\rm
    a}(1;1')\right\} = \log \int
\prod_{p=1}^{N}D{\bf r}^{(p)}(t)D{\bf \hat r}^{(p)}(t)
\exp
\Bigg\{\sum_{p=1}^{N}A_{0}[{\bf r}^{(p)},{\bf \hat r}^{(p)}]\nonumber\\ -
i\sum_{p=1}^{N}\int dt
r_{\alpha}^{(p)}(t)\left[\psi_{\alpha}\left({\bf r}^{(p)}(t)\right) +
  i\chi_{\alpha}\left({\bf r}^{(p)}(t)\right)\right]\nonumber\\ -
i\sum_{p=1}^{N}\int dt dt' p^{(p)}_{a}(t;t')\left[\Phi_{a}\left[{\bf
    r}^{(p)}(t);{\bf r}^{(p)}(t')\right] + iH_{a}\left[{\bf
    r}^{(p)}(t);{\bf r}^{(p)}(t')\right]\right]\Bigg\}\; . \label{61}
\end{eqnarray}
In order to ensure the extensivity of the whole effective action in
eq.(\ref{60}) we put the variance $\chi^2 = \chi_{0}^2 N$ (so that 
  the variance of the whole strength factor in eq.(\ref{3}) scaled as
  $N^{-1/2}$ akin to ref.\cite{23}) . This enable to represent GF in
  a similar to eq.(\ref{19}) form
\begin{eqnarray}
\left< Z \right>_{\rm av} \left\{\chi_{\alpha}\right\}=\int
\prod_{\alpha=0}^{1}\prod_{a=1}^{3}  D\psi_{\alpha}(1) D\Phi_{a}
\exp\Bigg\{ - N \widetilde{A}\left[\psi_{\alpha}, \Phi_{a} ; \chi_{\alpha}, H_{a}\right]\Bigg\}\; ,\label{62}
\end{eqnarray}
where
\begin{eqnarray}
 \widetilde{A}\left[\psi_{\alpha} ,\Phi_{a} ;\chi_{\alpha}, H_{a} \right]=
\frac{1}{2\xi_{0}}\int dt \int
d1 d2\psi_{\alpha}(1)\left[v^{-1}\right]_{\alpha\beta}(1,
2)\psi_{\beta}(2)\nonumber\\ 
  - \frac{1}{\chi_{0}^2}\int d1 d2 d3 d4 \Phi_{\rm
    a}(1;2)\left[\Gamma^{-1}\right]_{{\rm ab}}(1, 2, 3, 4) \Phi_{\rm
    b}(3;4)
-\frac{1}{N}\log \int
\prod_{p=1}^{N}D{\bf r}^{(p)}(t)D{\bf \hat r}^{(p)}(t)\nonumber\\
\times\exp
\Bigg\{\sum_{p=1}^{N}A_{0}[{\bf r}^{(p)},{\bf \hat r}^{(p)}] -
i\sum_{p=1}^{N}\int dt
r_{\alpha}^{(p)}(t)\left[\psi_{\alpha}\left({\bf r}^{(p)}(t)\right) +
  i\chi_{\alpha}\left({\bf r}^{(p)}(t)\right)\right] \nonumber\\
-i\sum_{p=1}^{N}\int dt dt' p^{(p)}_{a}(t;t')\left[\Phi_{a}\left[{\bf
    r}^{(p)}(t);{\bf r}^{(p)}(t')\right] + iH_{a}\left[{\bf
    r}^{(p)}(t);{\bf r}^{(p)}(t')\right]\right] \Bigg\}\; .\label{63}
\end{eqnarray}
The resulting SP - equation reads
\begin{eqnarray}
\overline{\psi_{\alpha}}(1)  &=&
-\frac{i\xi_{0}}{N}\int d2v_{\alpha\beta}(1, 2)\left<\rho_{\beta}(2)\right>_{{\rm SP}}\label{64}\\
\overline{\Phi_{a}}(1)  &=&
-\frac{i\xi_{0}}{N}\int d3 d4\Gamma_{ab}(1, 2, 3, 4)\left<\rho_{b}(2)\right>_{{\rm SP}}\; ,\label{SP}
\end{eqnarray}
where the average $<\cdots>_{SP}$ is calculated with the GF
\begin{eqnarray}
\left<Z_{0} \right>_{\rm av}\left\{\chi_{\alpha} , H_{a}\right\} &=&
\int
\prod_{p=1}^{N}D{\bf r}^{(p)}(t)D{\bf \hat r}^{(p)}(t)
\exp
\Bigg\{\sum_{p=1}^{N}A_{0}[{\bf r}^{(p)},{\bf \hat r}^{(p)}]\nonumber\\ 
 &-& i\sum_{p=1}^{N}\int dt
r_{\alpha}^{(p)}(t)\left[\overline{\psi_{\alpha}}\left({\bf r}^{(p)}(t)\right) +
  i\chi_{\alpha}\left({\bf r}^{(p)}(t)\right)\right]\nonumber\\ 
&-&i\sum_{p=1}^{N}\int dt dt' p^{(p)}_{a}(t;t')\left[\overline{\Phi_{a}}\left[{\bf
    r}^{(p)}(t);{\bf r}^{(p)}(t')\right] + iH_{a}\left[{\bf
    r}^{(p)}(t);{\bf r}^{(p)}(t')\right]\right]\Bigg\}\; . \label{65}
\end{eqnarray}
Thereby we are left with the GF of a free system which experiences the
external mean-fields $\overline{\psi_{\alpha}} + i\chi_{\alpha}$ and
$\overline{\Phi_{a}} +  iH_{a}$.

\subsection{The self-consistent Hartree approximation}
In order to calculate GF given by eq.(\ref{65}) we will use the
self-consistent Hartree approximation (SCHA). For this approximation we
replace the real action by an appropriate Gaussian one in such a way that all
terms which include more then two fields ${r}_{j}^{(p)}(t)$ or/and
$\hat{r}_{j}^{(p)}(t)$ are written in all possible ways as products of pairs
of ${r}_{j}^{(p)}(t)$ or $\hat{r}_{j}^{(p)}(t)$ coupled to self-consistent
averages of the remaining fields.

The analogy between SCHA and SP-approximation at $N \to \infty$ for the
special case when the non-quadratic terms in the action are only the functions
of the mean-squared displacement $d^2(t - t') = \sum_{p=1}^{N}\left<\left[{\bf
      r}^{(p)}(t) -{\bf r}^{(p)}(t')\right]^2\right>/N$ has been proven in
ref.\cite{8}.  In our case the action in eq.(\ref{65}) has a more general
form. In the Appendix E we show that the SCHA and the {\it next to the saddle
  point approximation} (NSPA) merge and both become exact, if the GF with an
arbitrary action can be treated by a steepest descent approach at $N \to
\infty$.  

Let us make the Fourier transformation of the mean-fields 
\begin{eqnarray}
\overline{\psi_{\alpha}}\left({\bf r}^{(p)}(t)\right) &=& \int
\frac{d^3 k}{(2\pi)^3}\overline{\psi_{\alpha}}\left({\bf k}\right)\exp\left\{i{\bf k} {\bf r}^{(p)}(t)\right\}\\\label{66}
\overline{\Phi_{a}}\left({\bf
    r}^{(p)}(t);{\bf r}^{(p)}(t')\right) &=& \int
\frac{d^3 k^1 d^3 k^2}{(2\pi)^6}\overline{\Phi_{a}}\left({\bf k^1},{\bf k^2}
\right)\exp\left\{i{\bf k^1} {\bf r}^{(p)}(t) +  \i{\bf k^2} {\bf r}^{(p)}(t')\right\}\label{67}
\end{eqnarray}
and insert it in eq.(\ref{65}) . Then for eq.(\ref{65}) we use the Hartree - type action 
(see eq.(\ref{140}). By doing so we put for simplicity the expectation
value $\xi_{0} = 0$ . It is easy to assure oneselves  also that the ``response
- response  overlap'' $\left<Q_{1}(1, 1')\right> = 0$ (similar to
$\left<\hat\sigma  \hat\sigma  \right> = 0$ in ref.\cite{26}). In the
curse of the derivation we have used SP - equation (\ref{SP})  and
defined the correlator (or the incoherent scattering function)
\begin{eqnarray}
C\left({\bf k^1}, t;{\bf k^2}, t'\right) = \frac{1}{N}\left<Q_{0}(
  {\bf k^1}, t;{\bf k^2}, t')\right>\label{68}
\end{eqnarray}
as well as the response functions
\begin{eqnarray}
G\left({\bf k^1}, t;{\bf k^2}, t'\right) &=& - \frac{1}{N}\left<Q_{3}(
  {\bf k^1}, t;{\bf k^2}, t')\right>\; \mbox{at}\; t' < t\\\label{69}
G\left({\bf k^2}, t;{\bf k^1}, t'\right) &=&  \frac{1}{N}\left<Q_{2}(
  {\bf k^1}, t;{\bf k^2}, t')\right>\; \mbox{at}\; t' >  t \; , \label{70}
\end{eqnarray}
where $\left<\cdots\right>$ stands for the averaging with the Hartree
- type of action. After collection of all terms the final result 
(at $\chi_{\alpha} = 0 \, \mbox{and} \,  H_{a} = 0$) reads then
\begin{eqnarray}
\left<Z_{0} \right>_{\rm av}\left\{\overline{\psi_{\alpha}} ,\overline{\Phi_{a}}\right\} &=&\int
\prod_{p=1}^{N}D{\bf r}^{(p)}(t)D{\bf \hat r}^{(p)}(t)
\exp
\Bigg\{\sum_{p=1}^{N}A_{0}[{\bf r}^{(p)},{\bf \hat r}^{(p)}] + \int d
t d t' i\hat{\bf r}^{(p)}(t) {\bf r}^{(p)}(t)\lambda(t ,
t')\nonumber\\
&-& \int d
t d t' i\hat{\bf r}^{(p)}(t) {\bf r}^{(p)}(t')\lambda(t ,
t') +  \int d
t d t' i\hat{\bf r}^{(p)}(t)i\hat{\bf r}^{(p)}(t')\eta(t ,
t')  \Bigg\}\; ,\label{71}
\end{eqnarray}
where 
\begin{eqnarray}
\lambda(t, t') = \frac{2}{3}\chi_{0}^2 \int\frac{d^3 k}{(2\pi)^3}
  k^2\left|v(k)\right|^2 G({\bf k}; t, t') C({\bf k}; t, t')\label{72}
\end{eqnarray}
and
\begin{eqnarray}
\eta(t, t') = \frac{1}{3}\chi_{0}^2 \int\frac{d^3 k}{(2\pi)^3}
  k^2\left|v(k)\right|^2 \left[ C({\bf k}; t, t')\right]^2 \; .\label{73}
\end{eqnarray}
In eqs.(\ref{71}) - (\ref{73}) we have restricted ourselves to the
homogeneous case
\begin{eqnarray}
 C({\bf k}, t;{\bf k'}, t') &=& (2\pi)^3 \delta({\bf k} + {\bf k'})
 C({\bf k}; t, t')  \nonumber\\
G({\bf k}, t;{\bf k'}, t') &=& (2\pi)^3 \delta({\bf k} + {\bf k'})
 G({\bf k}; t, t')
\end{eqnarray}
for the correlation and response function.
The equation of motion for the one particle correlator
\begin{eqnarray}
{\cal{P}}(t , t') = \frac{1}{3}\sum_{j=1}^{3}\left< r_{j}^{(p)}(t) r_{j}^{(p)}(t')\right>\label{74}
\end{eqnarray}
and the corresponding response function
\begin{eqnarray}
{\cal{G}}(t , t') = 
\frac{1}{3}\sum_{j=1}^{3}\left<i\hat r_{j}^{(p)}(t') r_{j}^{(p)}(t)\right>
\label{75}
\end{eqnarray}
(which actually does not depend from the particle index $p$) can be derived
from eq.(\ref{71}) by using the standard techniques \cite{8}. The resulting
equations read
\begin{eqnarray}
\left[m_{0}\frac{\partial^{2}}{\partial t^{2}} +
\gamma_{0}\frac{\partial}{\partial t} + \int\limits_{-\infty}^{t}d\tau 
 \lambda (t , \tau)\right]{\cal P}(t , t')\nonumber - 
\int\limits_{-\infty}^{t}d\tau 
 \lambda (t , \tau){\cal P}(t , t') \nonumber\\
+ \int\limits_{-\infty}^{t'}d\tau 
\: \eta (t , \tau){\cal G}(t', \tau) = -2T \gamma_{0}{\cal G}(t' , t)\label{76}
\end{eqnarray}
and
\begin{eqnarray}
\left[m_{0}\frac{\partial^{2}}{\partial t^{2}} +
\gamma_{0}\frac{\partial}{\partial t} + \int\limits_{-\infty}^{t}d\tau 
 \lambda (t , \tau)\right]{\cal G}(t , t')\nonumber - \int\limits_{-\infty}^{t}d\tau 
 \lambda (t , \tau){\cal G}(\tau , t') \nonumber\\
= - \delta(t - t')\; . \label{77}
\end{eqnarray}
Eqs.(\ref{76}) - (\ref{77}) should be supplemented with the initial
conditions $\gamma_{0}{\cal G}(t+0^{+},t) = -1$ and ${\cal G}(t,t) =
0$. By making use this condition , equipartition
$(m_{0}/3)\sum_{j=1}^{3}\left<{\dot r}_{j}(t){\dot
    r}_{j}(t)\right> = T$, causality ${\cal G}(t,t') = 0$ at $ t
\le t'$  as well as the condition $(1/3)\sum_{j=1}^{3}\left<{\dot
    r}_{j}(t) r_{j}(t)\right> = 0$ one find from eq.(\ref{76}) the
following equation
\begin{eqnarray}
\left[\frac{1}{2} m_{0}\frac{\partial^{2}}{\partial t^{2}}
  +\int\limits_{-\infty}^{t}d\tau \lambda (t , \tau)\right]{\cal P}(t,t)
 - \int\limits_{-\infty}^{t}d\tau  \lambda (t , \tau){\cal P}(\tau , t')
+ \int\limits_{-\infty}^{t}d\tau 
\: \eta (t , \tau){\cal G}(t, \tau) = 2T .\label{76'}
\end{eqnarray}

The set of eqs.(\ref{76}) - (\ref{77}) have the same structure as the
Dyson eq.(\ref{37}). After the matrices inversions and going to the
time domain eqs.(\ref{37}) (in time-translational invariant case) take 
the form
\begin{eqnarray}
\left[m_{0}\frac{\partial^{2}}{\partial t^{2}} +
\gamma_{0}\frac{\partial}{\partial t} + \mu(0)\right]G_{01}(t , t')
&-& \int\limits_{-\infty}^{t}d\tau \Sigma_{10}(t - \tau) G_{01}(\tau - t') =
\delta(t - t')\\\label{78}
\left[m_{0}\frac{\partial^{2}}{\partial t^{2}} +
\gamma_{0}\frac{\partial}{\partial t} + \mu(0)\right]G_{00}(t , t')
&-& \int\limits_{-\infty}^{t}d\tau \Sigma_{10}(t - \tau) G_{00}(\tau - t') \nonumber\\
&-&\int d\tau\Sigma_{11}(t - \tau) G_{10}(\tau - t') = 2T\gamma_{0} G_{10}(t 
- t')\; ,\label{79}
\end{eqnarray}
where $\mu(0) = \int\limits_{0}^{\infty}dt  \lambda (t)$ and
RPA - Fourier spectrum
\begin{eqnarray}
S_{01} = \frac{1}{-i\gamma_{0}\omega - m_{0}\omega^2 + \mu(0)}\; .\label{80}
\end{eqnarray}
Eqs.(\ref{76}) - (\ref{77}) are turned to the Dyson
equations (\ref{78}) - (\ref{79}) provided that
\begin{eqnarray}
G_{00}(t) &=& {\cal P} (t)  \quad,\quad  G_{01}(t) = - \: {\cal G} (t)\quad,\nonumber\\
\Sigma_{10}(t) &=& \lambda(t)  \quad,\quad  \Sigma_{11}(t) = \eta(t)\; .\label{81}
\end{eqnarray}
We can show \cite{17} that  the relation
\begin{eqnarray}
- \beta \frac{\partial}{\partial t} \Sigma_{11}(t) = \Sigma_{10}(t) -
\Sigma_{01}(t)
\label{82} 
\end{eqnarray}
holds,
provided that the FDT is satisfied for $G_{\alpha\beta}(t)$. We then have in addition
\begin{eqnarray}
- \beta \frac{\partial}{\partial t} G_{00}(t) = G_{01}(t) -
G_{10}(t) \; .
\label{83}
\end{eqnarray}
Bearing eqs.(\ref{81}) in mind the eq.(\ref{82}) takes in our case the 
form (at $t > 0$)
\begin{eqnarray}
- \beta\: \frac{\partial}{\partial t}\:\eta (t) = \lambda (t)\; .\label{84}
\end{eqnarray}
The validity of the relationship (\ref{84}) can be checked by
replacing (\ref{72}) and (\ref{73}) in (\ref{84}). 

The general eqs.(\ref{76}) - (\ref{77}) are equivalent, {\it mutatis
  mutandis}, to the corresponding equations for the $p$ - spin system or a
particle in the random potential at the large dimension \cite{6,7,8,8',9}.
The most important features of these equations are the {\it glassy dynamical
  behavior} and the universal {\it aging regime}. At low temperatures the
system tries to minimize the energy and each particle (with a number $p$)
tends to surround itself with other particles which assure the strength
parameter $\mu_{\rm pm} < 0$. On the other hand the long - range interaction tries
to support other pairs $(i j)$ corresponding to $\mu_{ij} > 0$. As a result
the system becomes ``frustrated'' and many local free energy minima appear.

In the spirit of ref.\cite{9,29,30} when $t, t' \to \infty$ we have to
discriminate between different cases: (i) the {\it asymptotic regime} when $(t
- t')/t \to 0$ and (ii) the {\it aging regime} when $(t - t')/t' \to {\cal
  O}(1)$. The aging regime is much more complicated because the time -
translational invariance and FDT are violated. This regime has been
extensively investigated both theoretically \cite{7,9,29,30} and by the
computer simulation \cite{31,32}. In the following  we restrict ourselves
only to the asymptotic regime, for the sake of clearness and  simplicity,
and since the main features will be already visible.

\subsection{The asymptotic regime}
This asymptotic regime is characterized by the large time scales, i.e.,
$t,t' \to \infty$ but keeping the difference $\tau = t -
t'$ finite. Under these circumstances we can define
\begin{eqnarray}
{\cal P}_{\rm as}(\tau) = \lim\limits_{t'\to \infty} {\cal P}(t' + \tau,
t')\nonumber\\
{\cal G}_{\rm as}(\tau) = \lim\limits_{t'\to \infty} {\cal G}(t' + \tau,
t')\label{85}
\end{eqnarray}
Then the equation for the displacement ${\cal D}_{\rm as} = 2 [{\cal
  P}_{\rm as}(0) - {\cal P}_{\rm as}(\tau)]$, response function ${\cal
  G}_{\rm as}(\tau)$ and the static correlator ${\cal P}_{\rm as}(0)$ takes
correspondingly the forms
\begin{eqnarray}
\left[m_{0}\frac{\partial^{2}}{\partial \tau^{2}} +
\gamma_{0}\frac{\partial}{\partial \tau} + M \right] {\cal D}_{\rm as}(\tau)
- \int\limits_{0}^{\tau}d\tau' \lambda_{\rm as} (\tau - \tau'){\cal
D}_{\rm as}(\tau') - \nonumber\\
-  \int\limits_{0}^{\infty}d\tau'\left[ \lambda_{\rm as}
(\tau + \tau') - \lambda_{\rm as}(\tau')\right]{\cal
D}_{\rm as}(\tau') - 2 \int\limits_{0}^{\infty}d\tau'\left[ \eta_{\rm as}
(\tau + \tau') - \eta_{\rm as}(\tau')\right]{\cal
G}_{\rm as}(\tau')= 2T \label{86}
\end{eqnarray}
\begin{eqnarray}
\left[m_{0}\frac{\partial^{2}}{\partial \tau^{2}} +
\gamma_{0}\frac{\partial}{\partial \tau} + M \right] {\cal G}_{\rm as}(\tau)
- \int\limits_{0}^{\tau}d\tau' \lambda_{\rm as} (\tau - \tau'){\cal
G}_{\rm as}(\tau') = 0\label{87}
\end{eqnarray}
\begin{eqnarray}
{\cal P}_{\rm as}(0) = \frac{1}{M - M_{\rm as}}\left[ T -
  \frac{1}{2}\int\limits_{0}^{\infty}d\tau \lambda_{\rm as}{\cal
      D}_{\rm as}(\tau) -\int\limits_{0}^{\infty}d\tau \eta_{\rm as}{\cal
      G}_{\rm as}(\tau)\right] ,\label{88} 
\end{eqnarray}
where
\begin{eqnarray}
M = \lim\limits_{t \to \infty}\int\limits_{-\infty}^{t}d\tau
\lambda(t, \tau)\; .
\label{89}
\end{eqnarray}
\begin{eqnarray}
M_{\rm as} = \int\limits_{0}^{\infty}d\tau \lambda_{\rm as}(\tau)
\; .
\label{90}
\end{eqnarray}
However, it is also convenient to define the ``anomaly'' ${\bar M} = M -
M_{\rm as}$ \cite{9}. The eqs.(\ref{86}) - (\ref{88}) has been analyzed
first in the context of polymeric manifold in the random media
\cite{8,8'} and the random - phase sine - Gordon model \cite{33}. The
peculiarity of our model is defined by its memory functions
$\lambda_{\rm as}(\tau)$ and $\eta_{\rm as}(\tau)$.

For example, let as give an explicit expression for $\eta_{\rm as}(\tau)$. The
Gaussian form  of the correlator, $C(\tau) = \exp\left\{ - k^2 {\cal
    D}_{\rm as}(\tau)/2\right\}$, leads from eq.(\ref{73})to the result
\begin{eqnarray}
\eta_{\rm as}(\tau) = \frac{\chi_{0}^2 \sqrt{\pi}}{6}\frac{1}{\sqrt{{\cal
      D}_{\rm as}(\tau)}}\; . \label{91}
\end{eqnarray}
Usually it is assumed that at the high temperature FDT holds, i.e.
\begin{eqnarray}
- \beta\: \frac{\partial}{\partial \tau}{\cal D}_{\rm as}(\tau) = 2 \:{\cal G}_{\rm as}(\tau)\label{92}
\end{eqnarray}
and
\begin{eqnarray}
- \beta\: \frac{\partial}{\partial \tau}{\eta}_{\rm as}(\tau) = 2 \:{\lambda}_{\rm as}(\tau)\; .\label{93}
\end{eqnarray}
In this case eqs.(\ref{86}) and  (\ref{87}) merge and take a simple form
\begin{eqnarray}
\left[m_{0}\frac{\partial^{2}}{\partial \tau^{2}} +
\gamma_{0}\frac{\partial}{\partial \tau} + M \right] {\cal D}_{\rm as}(\tau)
- \int\limits_{0}^{\tau}d\tau' \lambda_{\rm as} (\tau - \tau'){\cal
D}_{\rm as}(\tau') = 2T \; .\label{94}
\end{eqnarray}
It turns out \cite{8,8',32} that the solution which satisfies the FDT 
is only stable above a critical temperature $T_{\rm c}$. For the stability 
analysis it is convenient to represent eq.(\ref{94}) in the form
\begin{eqnarray}
\left[m_{0}\frac{\partial^{2}}{\partial \tau^{2}} +
\gamma_{0}\frac{\partial}{\partial \tau} +{\bar M} + M_{\rm as}(\tau) \right] {\cal D}_{\rm as}(\tau)
-\int\limits_{0}^{\tau}d\tau'\left[\eta_{\rm as} (\tau - \tau') - \eta_{\rm as}(\tau)\right]\frac{\partial}{\partial\tau'}{\cal
D}_{\rm as}(\tau') = 2T\; ,\label{95}
\end{eqnarray}
where 
\begin{eqnarray}
M_{\rm as}(\tau)  = \int\limits_{\tau}^{\infty}d\tau' \lambda_{\rm as}(\tau')\; .\label{96}
\end{eqnarray}
For $\tau \to \infty$ the stability condition which comes out of
eq.(\ref{95}) reads 
\begin{eqnarray} 
\left[ {\bar M} + M_{\rm as}(\tau)\right]{\cal
  D}_{\rm as}(\tau) \le 2 T \quad.\label{97}
\end{eqnarray}
Then the stationary value of the displacement ${\cal
  D}_{\rm as}(\tau \to \infty) =  q_{0}$ reads

\begin{eqnarray}
q_{0} = \frac{2 T}{{\bar M}}\quad.\label{98}
\end{eqnarray}
By taking into account eqs.(\ref{92}) and (\ref{93}) the stability
condition becomes
\begin{eqnarray}
D(q, T) \ge 0\label{99}
\end{eqnarray}
for $0 \le q \le q_{0}$, where 
\begin{eqnarray}
D(q, T) \equiv \left[ \left(\frac{\chi_{0}}{T}\right)^2
  \frac{\sqrt{\pi}}{12 \sqrt{q_{0}}} - \frac{1}{q_{0}}\right] q -
\left(\frac{\chi_{0}}{T}\right)^2 \frac{\sqrt{\pi}}{12} \sqrt{q} + 1 \; .\label{100}
\end{eqnarray}
The critical values $q_{\rm c}$ and $T_{\rm c}$ at which the condition
(\ref{99}) first becomes violated is defined by equations 
\begin{eqnarray}
D(q_{\rm c}, T_{\rm c}) &=& 0\nonumber\\ 
D'(q_{\rm c}, T_{\rm c}) &=& 0  \; .\label{102}
\end{eqnarray}
Consequently, eqs. (\ref{102}) have the simple solution
\begin{eqnarray}
\left(\frac{T_{\rm c}}{\chi_{0}}\right)^2 =  \frac{\sqrt{\pi q_{0}}}{24}
 \; \mbox{and}\;  q_{\rm c} = q_{0}\; .\label{103}
\end{eqnarray}

The Fig.3 shows the behavior of $D(q, T)$ in the vicinity of the
critical point. It can be seen that the minimum, $q_{m} \le q_{0}$, at
which $D(q, T) \le 0$ appears continuously, i.e. the instability of
the FDT solution shows up as a $2^{\rm nd}$ order phase transition. This
is analogous to the dynamics of polymeric manifolds in a medium with the
long range correlation in disorder \cite{8'}. In particular, if
``anomaly'' ${\bar M} \to 0$ then $q_{0} \to \infty$ and $T_{\rm c}
\to \infty$, so in this case the FDT solution is unstable for any
finite temperature.

\begin{center}
\begin{minipage}{12cm}
\begin{figure}
\epsfig{file=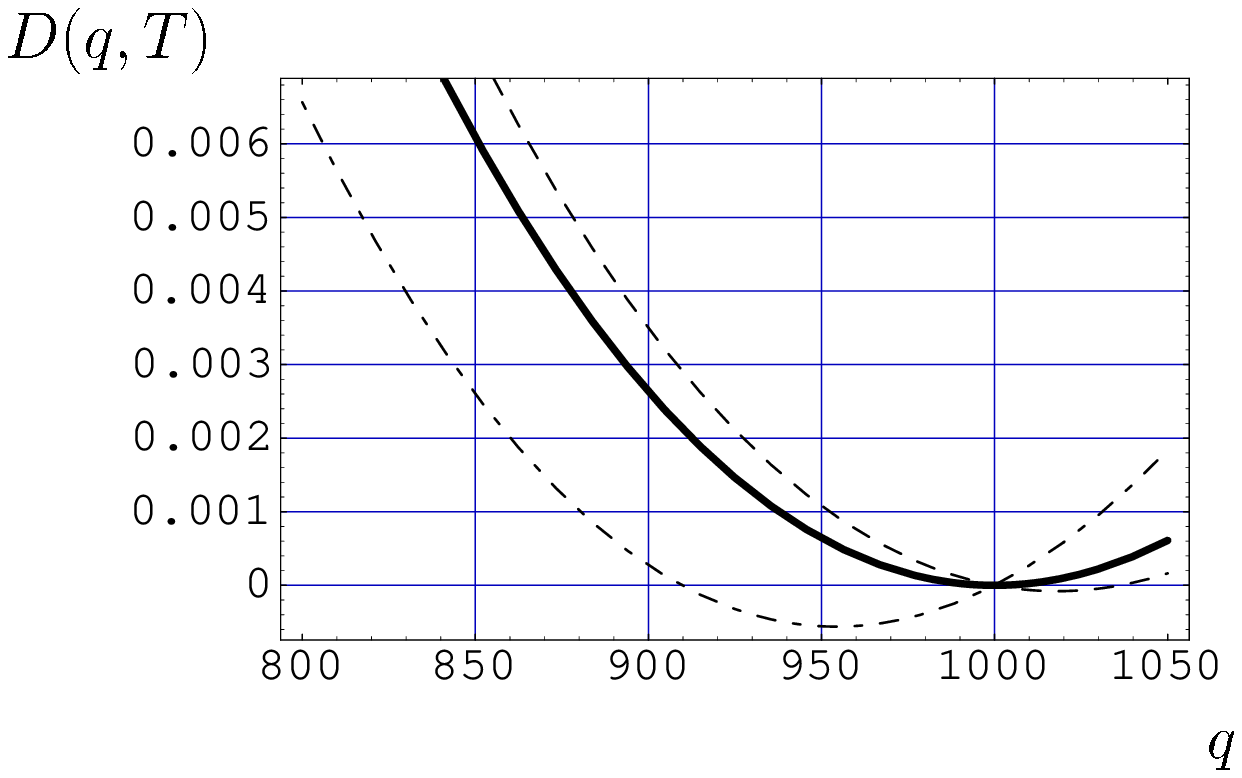,width=8cm}\\
{\caption 
{\footnotesize
$D(q,T)$ vs $q$ at $\chi_{0} = 0.1, q_{0} = 10^3$ for
  different temperatures: (i) full line corresponds to $T = T_{\rm c} =
  0.1528$; (ii) dashed  line $T = 0.1535$ ; (iii) dot - dashed  line $T = 0.151$}
}
\end{figure}
\end{minipage}
\end{center}

Let us consider the dynamics at the temperature slightly above the
critical point: $T = T_{\rm c}(1 + \varepsilon)$, where $0 < \varepsilon
\ll 1$. For large $\tau$ the decomposition
\begin{eqnarray}
{\cal D}_{\rm as}(\tau) = q_{0} - f(\tau)\; ,\label{104}
\end{eqnarray}
is possible, 
where $f(\tau) \ll q_{0}$. The substitution of this decomposition into 
eq.(\ref{95}) and the expansion up to the second order with respect to 
$f(\tau)$ yields
\begin{eqnarray}
\varepsilon q_{0}f(\tau) + \frac{1}{8}\left[f(\tau)\right]^2 +
\frac{1}{2}\int\limits_{0}^{\tau}d\tau'\left[f(\tau - \tau') -
  f(\tau)\right]\frac{\partial}{\partial\tau'}f(\tau') = 0 \; .\label{105}
\end{eqnarray}
Following ref.\cite{1} let us make the Laplace transformation ${\cal L}[f(\tau)] \equiv {\tilde f}(z)$
and introduce the scaling functions, ${\hat\phi}({\hat z})$ or ${\hat\phi}({\hat\tau})$ , in a such way
\begin{eqnarray}
{\tilde f}(z) =
\frac{c_{\varepsilon}}{\omega_{\varepsilon}}{\hat\phi}({\hat z})
\;\mbox{or} \; f(\tau) = c_{\varepsilon}{\hat\phi}({\hat\tau})\; ,\label{107} 
\end{eqnarray}
where ${\hat z} = z/\omega_{\varepsilon}$ and ${\hat\tau} =
\tau\omega_{\varepsilon}$. If $c_{\varepsilon} = \varepsilon$ and
$\omega_{\varepsilon} = \omega_{0} \varepsilon^{1/a}$ then one can
write eq.(\ref{105}) in the form
\begin{eqnarray}
q_{0}{\hat\phi}({\hat z}) - \frac{3}{8} {\cal
  L}\left\{{\hat\phi}^2({\hat\tau})\right\}({\hat z}) + {\hat
z}{\hat\phi}^2({\hat 
z}) = 0 \label{108}
\end{eqnarray}
(see eq.(2.68b) from ref.\cite{1}).

In the critical regime ${\hat z} \gg 1$ (or ${\hat\tau} \ll 1$) the
solution of eq.(\ref{108}) has a form ${\hat\phi}({\hat\tau}) \propto
{\hat\tau}^{-a}$. In this limit the first term in  eq.(\ref{108}) is
dropped out and the exponent is defined by the equation
\begin{eqnarray}
\frac{\Gamma^2 (1 - a)}{\Gamma (1 - 2a)} = \frac{3}{4}\; . \label{109}
\end{eqnarray}
The solution of eq. (\ref{109}) gives $a = 0.30465$. In the opposite
limit ${\hat z} \ll 1$ (or ${\hat\tau} \gg 1$) the last term in
eq.(\ref{108}) can be neglected. In this case the solution has the
form ${\hat\phi}({\hat\tau}) \propto
A_{\varepsilon}{\hat\tau}^{-a}\exp\{-\lambda {\hat\tau}\}$, where
$A_{\varepsilon} = 8 \varepsilon q_{0}\Gamma(1 - a)2^{(1 - 2a)}/
3\Gamma(1 - 2a)\lambda^2$. As a result the overall scaling reads 
\begin{eqnarray}
{\cal D}_{\rm as}(\tau) = \left\{
q_{0} -
  \frac{c_{\varepsilon}}{(\omega_{\varepsilon}\tau)^a} 
  ,\; \mbox{at}\; \omega_{\varepsilon}\tau \ll 1 \atop
 q_{0} - \frac{A_{\varepsilon}}{(\omega_{\varepsilon}\tau)^a}\exp\{ -
 \lambda (\omega_{\varepsilon}\tau)\} , \; \mbox{at}\;
 \omega_{\varepsilon}\tau \gg 1 \right.,\label{110}
\end{eqnarray}
where $\lambda$ is some constant.

At $T < T_{\rm c}$ the FDT is violated for the large time separation
$\tau$ and the aging regime is arising. It should be mentioned that
the asymptotic regime cannot be decoupled from the aging one
\cite{9,30}. In actual fact, the ``anomaly'' $\bar M$ in the asymptotic
eqs. (\ref{86}) - (\ref{88}) strictly speaking can  be calculated
only from the aging regime. Because of the distinct aim in this paper
we are not going to discuss the aging regime here expecting to return
to it in a later publication. 
 
\section{Conclusion}
In the present paper we have considered the dynamics of two models
with the long range repulsive interaction. The interaction potential
was designed in a way to enable the saddle point treatment, as well as a
fluctuation expansion. 

For the pure model we have derived eq.(\ref{48}) for the full correlation
matrix $G_{\alpha\beta}({\bf k}, \omega)$ in the one-loop approximation, which
has an the explicit solution (see eqs.(\ref{49'}) - (\ref{51})~). This
solution has a ``boring  behavior'' at $\omega \to 0$ which manifests the absence
of the glass dynamics. The physical background of this stems from the fact
that the potential is much too soft and the ``cage effect'' is completely
missing.

This conclusion is in accordance with the interacting particles
statistical thermodynamics analysis, which was given in
ref.\cite{10}. It was shown there that for the infinite range
interaction potential, which allows a well defined saddle point treatment, the
glassy phase is simply suppressed.

On the other hand, the same model but with a randomly distributed
strength of interaction (the ``random - bond  model'')  leads to the
continuous glass transition. This type of  transition is also the case 
for the polymeric manifolds in the disordered medium with long range
correlation \cite{8'} as well as for the $p$ - spin interaction
spin-glass model at the large external field \cite{3,5'}. It would be
also interesting to investigate the more realistic ``random sequence
model'' when each particle carry a random ``charge''.

Qualitatively the same glassy behavior have been found in the pure
spin  models with the deterministic but very rapidly oscillating
coupling between variables \cite{Bouchaud,Miglior,Krauth,Franz}. It
was assumed that the effective quenched disorder is ``self -
induced''\cite{7,Bouchaud}. This means that because of the slow
dynamics some degrees of freedom freeze and play the role of the
effectively quenched disorder.

As a conclusion, the glass transition in the pure systems of the
interacting particles, where the
disorder is actually ``self - induced'', goes beyond the
mean-field level \cite{10}. This appears to hard to implement in the present context,
because it implies the consideration of the short range interaction
potential as well as activated processes.

\section{Acknowledgments}
The authors have benefited from discussions with M. Fuchs, A. Latz, G.
Migliorini and  greatly indebted to  the Deutsche Forschungsgemeinschaft (DFG), the
Sonderforschungsbereich SFB 262 for financial support of the work.
\begin{appendix}
\section{Calculation of the 4-point RPA - correlation matrix}
In the full analogy with eq.(\ref{25}) the expression for
$S_{\alpha\beta\gamma\delta}^{(4)}(1,2,3,4)$ reads
\begin{eqnarray}
S_{\alpha\beta\gamma\delta}^{(4)}(1,2,3,4) = \lim_{\bar{\psi_{\alpha}}+ i\chi_{\alpha} \quad.
  \to 0}\left[ \frac{\delta}{N^2
    \delta\chi_{\beta}(2)\delta\chi_{\gamma}(3)\delta\chi_{\delta}(4)}
  \left< \rho_{\alpha}(1)\right>_{{\rm SP}}\right] \label{113}
\end{eqnarray}
The expansion of the $\left< \rho_{\alpha}(1)\right>_{{\rm SP}}$ up to 
the $3^{\rm d}$ order with respect to the mean field
$\bar{\psi_{\alpha}}+ i\chi_{\alpha}$ can be easily obtained from 
eq.(\ref{24}) 
\begin{eqnarray}
\left< \rho_{\alpha}(1)\right>_{{\rm SP}} &=& \left<
  \rho_{\alpha}(1)\right>_{0} + \int d2
\left<\Delta\rho_{\alpha}(1)\Delta\rho_{\beta}(2)\right>_{0} \left[
  \chi_{\beta}(2)  - i\bar{\psi_{\beta}}(2)\right]\nonumber\\
&+&\frac{1}{2!} \int
d2 d3\left<\Delta\rho_{\alpha}(1)\Delta\rho_{\beta}(2)\Delta\rho_{\gamma}(3)
\right>_{0} \left[\chi_{\beta}(2)  -
  i\bar{\psi_{\beta}}(2)\right] \left[\chi_{\gamma}(3)  -
  i\bar{\psi_{\gamma}}(3)\right]\nonumber\\
&+&\frac{1}{3!}\int
d2 d3 d4\left<\Delta\rho_{\alpha}(1)\Delta\rho_{\beta}(2)\Delta\rho_{\gamma}(3)
\Delta\rho_{\delta}(4) \right>_{0} \nonumber\\
&\times&\left[\chi_{\beta}(2)  -
  i\bar{\psi_{\beta}}(2)\right] \left[\chi_{\gamma}(3)  -
  i\bar{\psi_{\gamma}}(3)\right]\left[\chi_{\delta}(4)  -
  i\bar{\psi_{\delta}}(4)\right]\quad. \label{114} 
\end{eqnarray}
By using the SP-equation (\ref{23}) and after threefold
differentiation with respect to $\chi_{\alpha}(1)$ (see
eq.(\ref{113})) we find
\begin{eqnarray}
S_{\alpha\beta\gamma\delta}^{(4)}(1,2,3,4) &=&
F_{{\bar\alpha}{\bar\beta}{\bar\gamma}{\bar\delta}}^{(4)}({\bar 1},{\bar
  2},{\bar 3},{\bar 4})\nonumber\\
&\times&\left\{\left[{\hat 1} + \mu {\hat v} {\hat
      F}^{(2)}\right]^{-1} \right\}_{{\bar\alpha}\alpha}({\bar 1}, 1)\left\{\left[{\hat 1} + \mu {\hat v} {\hat
      F}^{(2)}\right]^{-1} \right\}_{{\bar\beta}\beta}({\bar 2},
2)\nonumber\\
&\times&\left\{\left[{\hat 1} + \mu {\hat v} {\hat
      F}^{(2)}\right]^{-1} \right\}_{{\bar\gamma}\gamma}({\bar 3}, 3)\left\{\left[{\hat 1} + \mu {\hat v} {\hat
      F}^{(2)}\right]^{-1} \right\}_{{\bar\delta}\delta}({\bar 4},
4)\quad, \label{115}
\end{eqnarray}
where the 4-point free system correlation matrix
\begin{eqnarray}
F_{\alpha\beta\gamma\delta}^{(4)}(1,2,3,4) = \frac{1}{N^2}\left<\Delta\rho_{\alpha}(1)\Delta\rho_{\beta}(2)\Delta\rho_{\gamma}(3)
\Delta\rho_{\delta}(4) \right>_{0}\quad. \label{116} 
\end{eqnarray}
In eq.(\ref{115}) we imply the summation (integration) over the barred
indices (barred space - time variables). When deriving
eq.(\ref{115}) we have also kept in mind that
$\left<\Delta\rho_{\alpha}(1)\Delta\rho_{\beta}(2)\right> \propto N$
and $\left<\Delta\rho_{\alpha}(1)\Delta\rho_{\beta}(2)\Delta\rho_{\gamma}(3)
\Delta\rho_{\delta}(4) \right>_{0} \propto N^2$ , etc. The fact that
the matrix $F_{\alpha\beta\gamma\delta}^{(4)}(1,2,3,4)$ is symmetrical 
with respect to simultaneous permutations of Greek indices and
space-time arguments as well as eq.(\ref{26}) have been used. 

It is easy to show that $F_{\alpha\beta\gamma\delta}^{(4)}(1,2,3,4)$
is factorized 
\begin{eqnarray}
F_{\alpha\beta\gamma\delta}^{(4)}(1,2,3,4) &=&
F_{\alpha\beta}^{(2)}(1,2) F_{\gamma\delta}^{(2)}(3,4) +
F_{\alpha\gamma}^{(2)}(1,3) F_{\beta\delta}^{(2)}(2,3) + F_{\alpha\delta}^{(2)}(1,4) F_{\beta\gamma}^{(2)}(2,3)\quad.
\end{eqnarray}
On the other side it is instructive to check that even in this case
$S_{\alpha\beta\gamma\delta}^{(4)}(1,2,3,4)$ can not be factorized.
\section{calculation of the vertex-matrix $K_{\alpha\beta}(1,2)$}
The substitution of the eg.(\ref{115}) into eq.(\ref{46}) after the
straightforward algebra yields
\begin{eqnarray}
K_{\alpha\beta}(1,2) &=& \left\{\left[\left(2 \mu \right)^{-1}{\hat v}^{-1} +
    {\hat F}\right]^{-1} - \left[ \mu^{-1} {\hat v}^{-1} +
    {\hat F}\right]^{-1}\right\}_{{\bar\beta}{\bar\alpha}}({\bar 2},
{\bar 1})\nonumber\\
&\times&\left\{F_{{\bar\alpha}{\bar\beta}}({\bar 1},{\bar
    2})F_{{\bar\gamma}{\bar\delta}}({\bar 3},{\bar 4}) + F_{{\bar\alpha}{\bar\gamma}}({\bar 1},{\bar
    3})F_{{\bar\beta}{\bar\delta}}({\bar 2},{\bar 4}) + F_{{\bar\alpha}{\bar\delta}}({\bar 1},{\bar
    4})F_{{\bar\beta}{\bar\gamma}}({\bar 2},{\bar
    3})\right\}\nonumber\\
&\times&\left\{\left[{\bar 1} + 2\mu {\bar v} {\bar
      F}\right]^{-1}\right\}_{{\bar\gamma}\alpha}({\bar 3}, 1)\left\{\left[{\bar 1} + 2\mu {\hat v} {\hat
      F}\right]^{-1}\right\}_{{\bar\delta}\beta}({\bar 4}, 2)\quad,
\end{eqnarray}
where as before for the repeated barred indices (variables) the
summation (integration) is implied. For the time - space -
translational invariant case the respective Fourier transformation
leads to the result:
\begin{eqnarray}
K_{\alpha\beta}({\bf k},\omega) &=& \Bigl\{ I
F_{{\bar\gamma}{\bar\delta}}({\bf k}, \omega) \nonumber\\
&+&\left\{\left[\left(2 \mu \right)^{-1}{\hat v}^{-1} +
    {\hat F}\right]^{-1} - \left[\mu^{-1}{\hat v}^{-1} +
    {\hat F}\right]^{-1}\right\}_{{\bar\beta}{\bar\alpha}}(-{\bf k}, -\omega)
 F_{{\bar\alpha}{\bar\gamma}}(-{\bf
   k},-\omega)F_{{\bar\beta}{\bar\delta}}({\bf k},\omega)\nonumber\\
&+&\left\{\left[\left(2 \mu \right)^{-1}{\hat v}^{-1} +
    {\hat F}\right]^{-1} - \left[\mu^{-1}{\hat v}^{-1} +
    {\hat F}\right]^{-1}\right\}_{{\bar\beta}{\bar\alpha}}({\bf k}, \omega)
 F_{{\bar\alpha}{\bar\delta}}({\bf
   k},\omega)F_{{\bar\beta}{\bar\gamma}}(-{\bf k},-\omega)\Bigr\}\nonumber\\
&\times&\left\{\left[{\bar 1} + 2\mu {\hat v} {\hat
      F}\right]^{-1}\right\}_{{\bar\gamma}\alpha}(-{\bf k}, -\omega)\left\{\left[{\bar 1} + 2\mu {\bar v} {\bar
      F}\right]^{-1}\right\}_{{\bar\delta}\beta}({\bf k}, \omega)\quad,
\end{eqnarray}
where the trace
\begin{eqnarray}
I = \int \frac{d^{3}q d\omega}{(2\pi)^4}\left\{\left[{\hat 1} + \left(
      2 \mu \right)^{-1}{\hat F}^{-1}{\hat v}^{-1}\right]^{-1} -
  \left[{\hat 1} + \mu^{-1}{\hat F}^{-1}{\hat v}^{-1}\right]^{-1}
\right\}_{{\bar\alpha}{\bar\alpha}}({\bf q}, \omega)\quad.
\end{eqnarray}
With the
correlation matrix ${\hat F}$ given by eqs.(\ref{49}),  by doing
integration over $\omega$ one can check that the trace $I = 0$. This
gives finally
\begin{eqnarray}
K_{\alpha\beta}({\bf k}, \omega) = L_{\alpha\beta}({\bf k}, \omega) + L_{\alpha\beta}({-\bf k}, -\omega)\quad,
\end{eqnarray}
where
\begin{eqnarray}
L_{\alpha\beta}({\bf k}, \omega) &=& \left\{\left[{\bar 1} + 2\mu {\hat v} {\hat
      F}\right]^{-1}\right\}_{\alpha{\bar\gamma}}({\bf k},
\omega)F_{{\bar\gamma}{\bar\beta}}({\bf k},\omega)\nonumber\\
&\times&\left\{\left[{\hat 1} + \left(
      2 \mu \right)^{-1}{\hat F}^{-1}{\hat v}^{-1}\right]^{-1} -
  \left[{\hat 1} + \mu^{-1}{\hat F}^{-1}{\hat v}^{-1}\right]^{-1}
\right\}_{{\bar\beta}{\bar\delta}}({\bf q}, \omega)\nonumber\\
&\times&\left\{\left[{\bar 1} + 2\mu {\hat v} {\hat
      F}\right]^{-1}\right\}_{\alpha{\bar\gamma}}({\bf k},
\omega)\quad .
\end{eqnarray}
\section{The MCT for the generalized Kac potential}
 In this case the direct correlation function $c({\bf
  r}) = - \beta V({\bf r})$ and its Fourier transformation takes the scaling
form
\begin{eqnarray}
c({\bf k}) = - \beta f\left(\frac{{\bf k}}{\kappa}\right)\; .\label{Corr}
\end{eqnarray}
Let us insert this expression to the MCT - memory kernel (see
eq.(3.32) in \cite{1}). It is reasonable then to rescale the
integration variables in the memory kernel , ${\bf k} \to \kappa {\bf
  k},\;  {\bf p} \to \kappa {\bf p}$ as well as to put for the external
wave vector ${\bf q} = \kappa {\bf q}_{0}$ , where ${\bf q}_{0}$ is
some reference wave vector. The last scaling means that in the MF -
limit an experiment probes a very small wave vector : $ {\bf q} \to
0$. The resulting scaling of the memory kernel , $m({\bf q}, t)$, reads
\begin{eqnarray}
m(\kappa {\bf q}_{0}, t) &=& \kappa^d \widetilde{S}({\bf
  q}_{0})\:\frac{\rho_{0}}{2}\: \int \frac{d{\bf k}d{\bf p}}{(2\pi)^{2d}}
\:\delta^{(d)} ({\bf k} + {\bf p} - {\bf q}_{0}) \nonumber\\
&\times&\frac{\left\{e^{L}({\bf q})\beta \left[{\bf k} f({\bf k}) + {\bf p}
    f({\bf p})\right]\right\}^2}{q_{0}^2}\: \widetilde{S}({\bf k}, t)
\:\widetilde{S}({\bf p}, t)\; ,\label{Kernel}
\end{eqnarray}
where we have took into account  the scaling form of the correlator:
$S({\bf k} , t) = \widetilde{S}({\bf k}/\kappa ; t)$. Thus we finally
arrive at the scaling $m(\kappa {\bf q}_{0}, t) \propto \kappa^d \to
0$ and the glass transition dies out.
\section{The analogy between  SCHA and  NSPA}
Let us prove that SCHA becomes exact for GF given by  (\ref{65}) in
the limit $N \to \infty$. We will consider even a  more general GF
\begin{eqnarray}
Z\{\chi_{\alpha}\} &=& \int\prod_{p=1}^{N}\prod_{\alpha=0, 1}Dx_{\alpha}^{(p)}(1)\:\exp
\Bigg\{-\frac{1}{2}\sum_{p=1}^{N}\int d1 d2
\:x_{\alpha}^{(p)}(1)A_{\alpha\beta}(1, 2)x_{\beta}^{(p)}(2) \nonumber\\ 
 &+&\sum_{p=1}^{N}W\left[x_{\alpha}^{(p)}\right] + \sum_{p=1}^{N}\int
 d1 x_{\alpha}^{(p)}(1)\chi_{\alpha}(1)\Bigg\} \;,\label{117}
\end{eqnarray}
where we have used the shorthand notations
\begin{eqnarray}
x_{\alpha}^{(p)}(1) = \left(\begin{array}{c}
r_{j}(t)\\i{\hat r}_{j}(t)\label{118}
\end{array}\right)\; ,
\end{eqnarray}
and ``1'' embraces Cartesian indices  as well as time variable: $1
\equiv \{i, j, k; t\}$. In eq.(\ref{117})
$W\left[x_{\alpha}^{(p)}\right]$ is an arbitrary functional of
$x_{\alpha}^{(p)}$.

Instead of the exact action functional in eq.(\ref{117}) we consider
now the trial one which has a Gaussian form
\begin{eqnarray}
S\left[x_{\alpha}^{(p)}(1) \right] &=& \sum_{p=1}^{N}\Bigg\{ \frac{1}{2} \int d1 d2
\:x_{\alpha}^{(p)}(1)A_{\alpha\beta}(1,
2)x_{\beta}^{(p)}(2)\nonumber\\
&-&\int d1 d2\:x_{\alpha}^{(p)}(1){\Gamma}_{\alpha\beta}(1,
2)x_{\beta}^{(p)}(2) - \int d1 L_{\alpha}(1)x_{\alpha}^{(p)}(1)\Bigg\} 
\; .\label{119}
\end{eqnarray}
Let us look for the ``best'' coefficients ${\Gamma}_{\alpha\beta}(1,
2)$ and $L_{\alpha}(1)$ in a sense that the exact ``free energy''
$F\left[\chi_{\alpha}\right] = - \log Z\left\{\chi_{\alpha}\right\}$
tends to the trial one $F_{0}\left[\chi_{\alpha}\right] = - \log
\int\prod D x_{\alpha}^{(p)}\exp \{ - S[ x_{\alpha}^{(p)}]\}$,
  i. e.
\begin{eqnarray}
F\left[\chi_{\alpha}\right] \longrightarrow F_{0}\left[\chi_{\alpha}\right]\label{120}
\end{eqnarray}
and both becomes exact at $N \to \infty$.

We can show that the property (\ref{119}) is satisfied by
$\Gamma_{\alpha\beta}$ and $L_{\alpha}$ which are obtained by
extremization of the functional
\begin{eqnarray}
\Phi\left\{\Gamma_{\alpha\beta},L_{\alpha}\right\} &=& - \log
\int\prod_{p=1}^{N}\prod_{\alpha=0,
  1}Dx_{\alpha}^{(p)}(1)\:\exp\Bigg\{ - S\left[
  x_{\alpha}^{(p)}\right]\Bigg\} \nonumber\\
&+& \sum_{p=1}^{N}\Bigg\{\int d1 d2 \Gamma_{\alpha\beta}(1, 2)
\left<x_{\alpha}^{(p)}(1)x_{\beta}^{(p)}(2)\right>_{\rm s}\nonumber\\ 
&+& \int d1
\left[L_{\alpha}(1) - \chi_{\alpha} \right]
\left<x_{\alpha}^{(p)}(1)\right>_{\rm s} -
\left<W\left[x_{\alpha}^{(p)}(1)\right]\right>_{\rm s} \Bigg\}\; ,\label{121}
\end{eqnarray}
where we use the notations
\begin{eqnarray}
\left<\dots\right>_{\rm s} = \frac{\int\prod_{p=1}^{N}\prod_{\alpha=0,
  1}Dx_{\alpha}^{(p)}(1)\: \dots \exp\Bigl\{ - S\left[
  x_{\alpha}^{(p)}\right]\Bigr\}}{\int\prod_{p=1}^{N}\prod_{\alpha=0,
  1}Dx_{\alpha}^{(p)}(1)\:\exp\Bigl\{ - S\left[
  x_{\alpha}^{(p)}\right]\Bigr\}} \;.\label{122}
\end{eqnarray}
The extremization conditions reads
\begin{eqnarray}
\frac{\delta}{\delta \Gamma_{\alpha\gamma}(1,2)}\:  \Phi &=& 0 \nonumber\\
\frac{\delta}{\delta L_{\alpha}(1)}\: \Phi &=& 0\;.\label{123}
\end{eqnarray}
The variations in eqs.(\ref{122}) can be done directly. During the
calculation the generalized Wick's theorem \cite{20} should be also
taken into account. Namely, because the averaging (\ref{121}) is
simply the Gaussian integral the Wick's theorem yields 
\begin{eqnarray}
\left<x_{\alpha}^{(p)}(1)W\left[x_{\alpha}^{(p)}\right]\right>_{\rm s} 
&=& \left<x_{\alpha}^{(p)}(1)\right>_{\rm s}\left< W\left[x_{\alpha}^{(p)}\right]\right>_{\rm s} \nonumber\\
&+&\int d2 \:\left<\Delta x_{\alpha}^{(p)}(1)\Delta x_{\beta}^{(p)}(2)
\right>_{\rm s}\:\left<\frac{\delta}{\delta x_{\beta}^{(p)}(2)}
 \: W\left[x_{\alpha}^{(p)}\right]\right>_{\rm s}\; ,\label{124}
\end{eqnarray}
where $\Delta x_{\alpha}^{(p)}(1) \equiv x_{\alpha}^{(p)}(1) - \left<
  x_{\alpha}^{(p)}(1)\right>_{\rm s}$. After the straightforward
calculation we find
\begin{eqnarray}
\Gamma_{\alpha\gamma}(1,2) = \frac{1}{2}\left<\frac{\delta^2}{\delta x_{\alpha}^{(p)}(1) \delta x_{\beta}^{(p)}(2)}
 \: W\left[x_{\alpha}^{(p)}\right]\right>_{\rm s}\label{125}
\end{eqnarray}
and
\begin{eqnarray}
L_{\alpha}(1) = \int d2 \left[ A_{\alpha\beta}(1, 2) -
  \Gamma_{\alpha\beta}(1, 2)\right]\:\left<
  x_{\alpha}^{(p)}(2)\right>_{\rm s}\; .\label{126}
\end{eqnarray}
Then equations for the two moments take the form
\begin{eqnarray}
\left<
  x_{\alpha}^{(p)}(1)\right>_{\rm s} = \int d2 \left[
  A^{-1}\right]_{\alpha\beta]}(1, 2) \left[\left<\frac{\delta}{\delta
        x_{\beta}^{(p)}(2)} W\left[x_{\alpha}^{(p)}\right]\right>_{\rm
      s} + \chi_{\beta}(2)\right]\label{127}
\end{eqnarray}
and
\begin{eqnarray}
\left<\Delta x_{\alpha}^{(p)}(1)\Delta x_{\beta}^{(p)}(2)
\right>_{\rm s} = \left\{\left[ {\hat A} - 2 \: {\hat
      \Gamma}\right]^{-1}\right\}_{\alpha\beta}(1, 2) \; ,\label{128}
\end{eqnarray}
where ${\hat A}$ and ${\hat \Gamma}$ stands for the corresponding $2
\times 2$ - matrices.

On the other side the saddle point (SP) treatment of eq.(\ref{117}) at $N
\to \infty$ yields
\begin{eqnarray}
-\int d2 \: A_{\alpha\beta}(1, 2)\: \bar{x}_{\beta}^{(p)}(2) + \left.
\frac{\delta \:W}{\delta x_{\alpha}^{(p)}(1)}\: \right|_{x_{\alpha} =
\bar{x}_{\alpha}} + \chi_{\alpha}(1) = 0 \label{129}
\end{eqnarray}
and
\begin{eqnarray}
\left<\Delta x_{\alpha}^{(p)}(1)\Delta x_{\beta}^{(p)}(2)
\right>_{\rm SP}  = \left\{\left[ {\hat A} - 2 \: {\hat B}\right]^{-1}\right\}_{\alpha\beta}(1, 2) \; ,\label{130}
\end{eqnarray}
where 
\begin{eqnarray}
 B_{\alpha\beta}(1, 2) = \frac{1}{2}\: \left.\frac{\delta^2 \: W}{\delta
     x_{\alpha}^{(p)}(1)\delta x_{\beta}^{(p)}(2)}\: \right|_{x_{\alpha} =
\bar{x}_{\alpha}}\label{131}
\end{eqnarray}
and $\bar{x}_{\alpha}^{(p)}(1)$ stands for the field in SP.

In order to show the analogy between eqs.(\ref{127})-(\ref{128}) and
eqs.(\ref{129})-(\ref{130}) let us make the functional Fourier transformation
\begin{eqnarray}
\exp \left\{K\left[ y_{\alpha}^{(p)}(1)\right]\right\} = \int D
x_{\alpha}^{(p)}(1)\exp\left\{W\left[x_{\alpha}^{(p)}(1)\right] - i\int d1\:x_{\alpha}^{(p)}(1)y_{\alpha}^{(p)}(1)\right\}\label{132}
\end{eqnarray}
and its inversion
\begin{eqnarray}
\exp \left\{W\left[ x_{\alpha}^{(p)}(1)\right]\right\} = \int D
y_{\alpha}^{(p)}(1)\exp\left\{K\left[y_{\alpha}^{(p)}(1)\right] + i\int d1\:x_{\alpha}^{(p)}(1)y_{\alpha}^{(p)}(1)\right\}\; .\label{133}
\end{eqnarray}
Then eqs.(\ref{129}) - (\ref{130}) can be written as 
\begin{eqnarray}
\bar{x}_{\alpha}^{(p)}(1) = \int
d2\left[A^{-1}\right]_{\alpha\beta}(1,
2)\left[i\left<y_{\beta}^{(p)}(2)\right>_{\rm SP} + \chi_{\beta}(2)\right]\label{134}
\end{eqnarray}
and
\begin{eqnarray}
\left<\Delta x_{\alpha}^{(p)}(1)\Delta x_{\beta}^{(p)}(2)
\right>_{\rm SP}  = \left\{\left[ {\hat A} + \Vert \left<\Delta y
      \Delta y\right>_{\rm SP} \Vert
  \right]^{-1}\right\}_{\alpha\beta}(1, 2)\; ,\label{135}
\end{eqnarray}
where the correlation matrix 
\begin{eqnarray}
\Vert \left<\Delta y\Delta y\right>_{\rm SP} \Vert = \left<\Delta
  y_{\alpha}^{(p)}(1)\Delta y_{\beta}^{(p)}(2)\right>_{\rm SP}\label{136}
\end{eqnarray}
and 
\begin{eqnarray}
\left< \dots \right>_{\rm SP} = \frac{\int D y_{\alpha}^{(p)} \dots
  \exp\left\{K\left[y_{\alpha}^{(p)}\right] + i\int d1\:\bar{x}_{\alpha}^{(p)}(1)y_{\alpha}^{(p)}(1)\right\}}{\int D y_{\alpha}^{(p)}
  \exp\left\{K\left[y_{\alpha}^{(p)}\right] + i\int d1\:\bar{x}_{\alpha}^{(p)}(1)y_{\alpha}^{(p)}(1)\right\}}\; .\label{137}
\end{eqnarray}
At $N \to \infty $ by making use eq.(\ref{133}) one can immediately
see that
\begin{eqnarray}
\left<\frac{\delta}{\delta
        x_{\alpha}^{(p)}(1)} W\left[x_{\alpha}^{(p)}\right]\right>_{\rm 
      s} \longrightarrow i\left< y_{\alpha}^{(p)}(1)\right>_{\rm SP}\label{138}
\end{eqnarray}
and
\begin{eqnarray}
\left<\frac{\delta^2}{\delta x_{\alpha}^{(p)}(1) \delta x_{\beta}^{(p)}(2)}
 \: W\left[x_{\alpha}^{(p)}\right]\right>_{\rm s} \longrightarrow
 - \left<\Delta  y_{\alpha}^{(p)}(1)\Delta  y_{\beta}^{(p)}(2) \right>_{\rm SP}\; ,\label{139}
\end{eqnarray}
and SCHA exactly corresponds to NSPA . For the case which was treated
in Sec.III C $\left< x_{\alpha}^{(p)}(1)\right>_{\rm s} = 0$ and the
Hartree - type action (\ref{119}) cast the form 
\begin{eqnarray}
S\left[x_{\alpha}^{(p)}(1) \right] &=& \sum_{p=1}^{N}\Bigg\{ \frac{1}{2} \int d1 d2
\:x_{\alpha}^{(p)}(1)A_{\alpha\beta}(1,
2)x_{\beta}^{(p)}(2)\nonumber\\
&-&\frac{1}{2}\int d1 d2\:\left<\frac{\delta^2}{\delta x_{\alpha}^{(p)}(1) \delta x_{\beta}^{(p)}(2)}
 \: W\left[x_{\alpha}^{(p)}\right]\right>_{\rm s}  x_{\alpha}^{(p)}(1)  x_{\beta}^{(p)}(2) \Bigg\} 
\; .\label{140}
\end{eqnarray}

\end{appendix}

\newpage
\begin{figure}
\epsfig{file=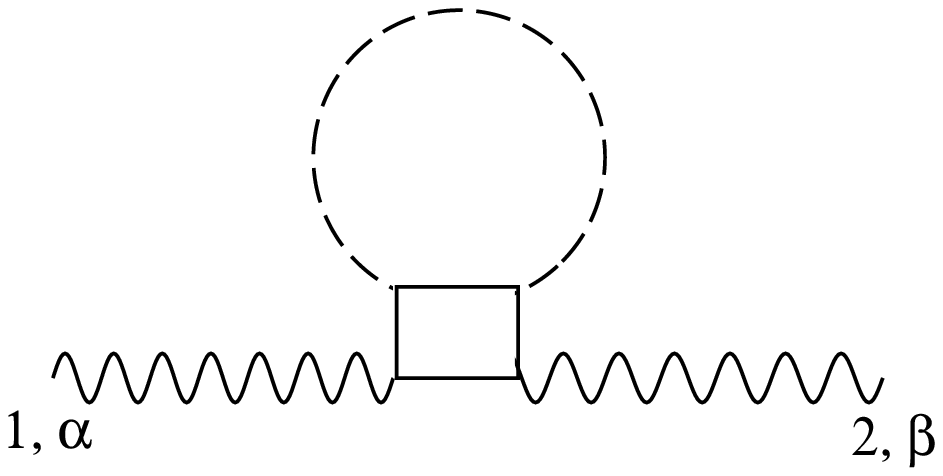,width=12cm, height=6cm}
\vspace{2cm}
\caption{Diagramatic interpretation of the vertex - matrix: the
  rectangle corresponds to
  $S_{\alpha\beta\gamma\delta}^{(4)}(1,2,3,4)$; the dash line - to the 
  effective interaction matrix $\left[(\mu {\hat v})^{-1} + {\hat
      S}\right]^{-1}$; the wave line - to $\left[ {\hat 1} + \mu {\hat 
      v}{\hat S}\right]^{-1}$.}
\end{figure}
\newpage
\vspace{4cm}
\begin{figure}
\epsfig{file=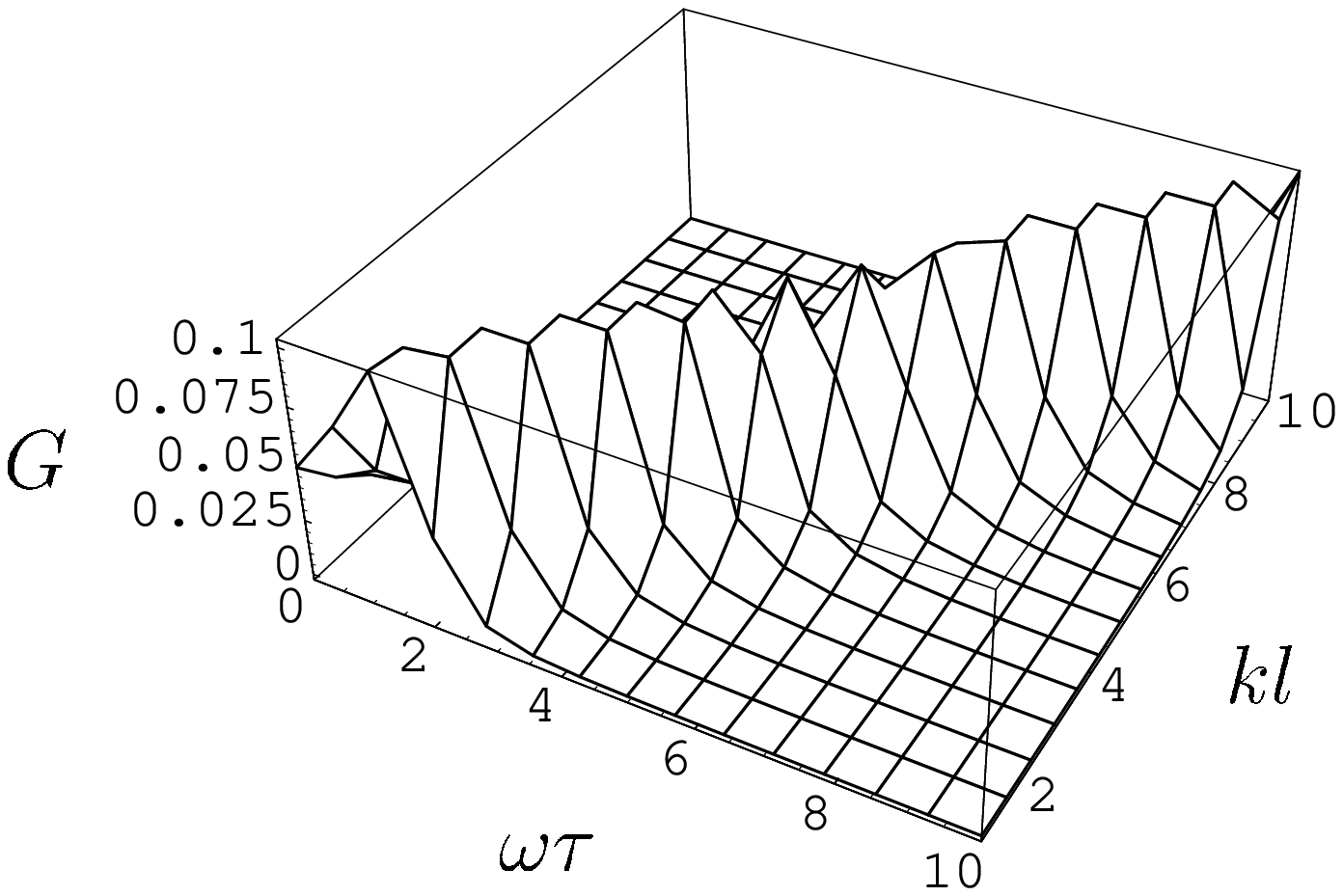,width=12cm, height=9cm}
\vspace{2cm}
\caption{The correlation function $G_{00}({\bf k},\omega)$ vs rescaled 
  variables $\omega\tau_{0}$  and  $kl_{0}$, where $\tau_{0} =
  m_{0}/\gamma_{0}$ and $l_{0} = (\tau_{0}/\beta\gamma_{0})^{1/2}$}
\end{figure}
\newpage
\begin{figure}
\epsfig{file=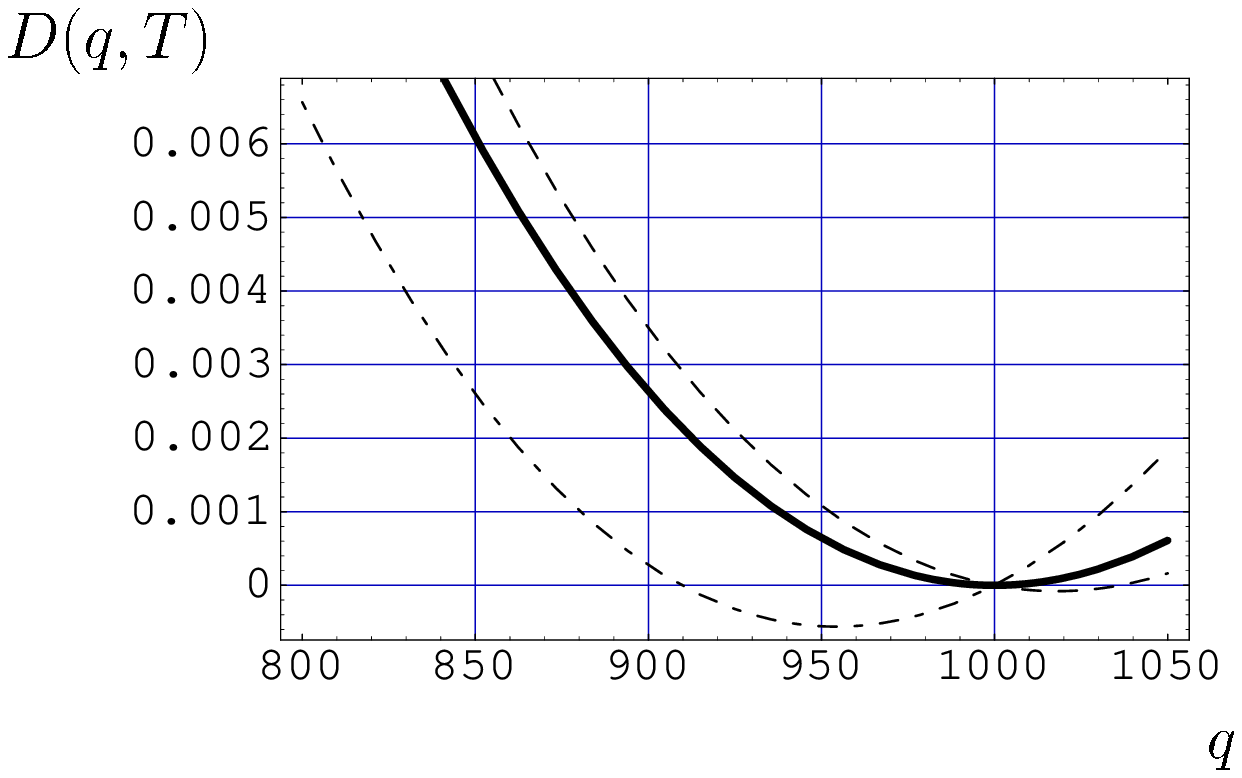,width=13.5cm, height=8.5cm}
\vspace{2cm}
\caption{$D(q,T)$ vs $q$ at $\chi_{0} = 0.1, q_{0} = 10^3$ for
  different temperatures: (i) full line corresponds to $T = T_{\rm c} =
  0.1528$; (ii) dashed  line $T = 0.1535$ ; (iii) dot - dashed  line $T = 0.151$}
\end{figure}

\end{document}